\documentstyle[12pt]{article}

\newcommand{\be}{\begin{equation}}
\newcommand{\ee}{\end{equation}}
\newcommand{\nn}{\noindent}
\newcommand{\beba}{\begin{equation}\begin{array}{lcl}}
\newcommand{\eaee}{\end{array}\end{equation}}
\newcommand{\bea}{\begin{eqnarray}}
\newcommand{\eea}{\end{eqnarray}}
\newcommand{\ba}{\begin{array}}
\newcommand{\ea}{\end{array}}

\newcommand{\dt}{\mathaccent95}
\newcommand{\nnu}{\nonumber}
\def\chib{{\mbox{\boldmath $\chi$}}}

\newcommand{\ds}{\displaystyle}
\newcommand{\td}{\tilde}

\newcommand{\sis}{\scriptscriptstyle}%
\newcommand{\norsl}{\normalsize\sl}
\newcommand{\norsc}{\normalsize\sc}
\def\a{\alpha}
\def\b{\beta}
\def\g{\gamma}
\def\c{\chi}
\def\d{\delta}
\def\e{\epsilon}
\def\ep{\varepsilon}

\def\m{\mu}
\def\n{\nu}

\def\o{\omega}
\def\p{\pi}

\def\r{\rho}
\def\s{\sigma}

\def\x{\xi}

\def\cd{{\cal D}}

\def\ch{{\cal H}}

\def\cl{{\cal L}}

\begin{document}

\textheight 22cm
\voffset  -1cm
\begin{titlepage}

\vskip 4cm

\title{\vskip 1.4cm \Large\bf { Quantum Cosmology with Yang-Mills Fields }}

\author{
\norsc  D. Kapetanakis$^1$, G. Koutsoumbas$^2$,
A. Lukas$^3$ and P. Mayr$^3$  \\
\norsl $^1$ Laboratoire de Physique Th\'eorique\\
\norsl Chemin de Bellevue, F-74941\\
\norsl Annecy-le-Vieux Cedex, France\\
\\
\norsl  $^2$Physics Department\\
\norsl  National Technical University\\
\norsl  157 80 Zografou, Athens, Greece\\
\\
\norsl  $^3$Physik Department\\
\norsl  Technische  Universit\"{a}t M\"{u}nchen\\
\norsl  D-85747 Garching, Germany\\     }

\date{}
\maketitle

\begin{abstract}
{\normalsize \noindent We examine an extension of
the ideas of quantum cosmology and, in
particular, the proposal of Hartle and Hawking for the boundary conditions of
the Universe, to models which incorporate Yang-Mills fields. Inhomogeneous
perturbations about a homogeneous, isotropic minisuperspace background model
are considered, by expanding the Yang-Mills fields in harmonics of the spatial
directions which are
taken to be three-spheres. The expansions are made explicit for
$SO(N)$ gauge fields thereby obtaining formulae compatible with the formalism
conventionally used in quantum cosmology. We apply these results to the gauge
group $SO(3)$ and derive the Lagrangian and the semi-classical wave function
for this special case.}
\end{abstract}

\begin{picture}(5,2)(-300,-690)
\put(2.3,-100){ENSLAPP-A-434/93}
\put(2.3,-120){TUM--TH--160/93}
\put(2.3,-140){NTUA 43/93}
\put(2.3,-160){hep-th/9403131}
\put(2.3,-190){March 1994}
\end{picture}

\thispagestyle{empty}
\end{titlepage}
\setcounter{page}{1}
\baselineskip=6mm
\renewcommand{\arraystretch}{1.7}
\section*{1. Introduction}
Recently we have witnessed a great activity in the interplay between high
energy physics and cosmology. The question of the initial conditions which in
the early universe gave origin to the universe that we observe is clearly one
of the main questions asking for clarification. Obviously, progress in such a
direction needs an understanding of quantum gravity which is still lacking.
On the other hand quantum cosmology provides a useful setting  to
discuss initial conditions and quantum gravity ideas. The most interesting
results in this area were obtained from the so--called
minisuperspace models where the infinite number of modes of the gravitational
and matter fields was reduced by freezing out all the modes except of a finite
number which  are assumed to dominate . The quantization then proceeds
through the application of the ADM~\cite{adm} formalism, thus obtaining the
Wheeler-DeWitt equation as the Hamiltonian constraint and interpreting the
conjugate momenta as operators, plus momentum constraints. It is clear that
in order to go beyond the minisuperspace analysis one has to include the
infinite number of the field modes, treated as frozen previously.
This is usually done by treating these modes as perturbations around
the minisuperspace degrees of freedom.

In this context, Hartle and Hawking~\cite{hah,hh}  worked out
the no-boundary proposal. An alternative suggestion is the so--called
tunneling boundary condition~\cite{tun}.
Excellent reviews already exist on the subject~\cite{reviews}.
However, in most of the minisuperspace models and their extensions including
perturbations which exist so far, the matter source is taken to be
a scalar field~\cite{hh}.
Also fermion fields~\cite{dh} and electromagnetism~\cite{jl} have been
considered.  Though this list
contains fundamental fields, the absence of non--Abelian gauge fields is
striking, especially if one considers that the evolution of the very early
universe is dominated by gravitational and non--Abelian gauge fields.
Recently a minisuperspace model including gauge
fields was constructed~\cite{bmpv} and the wave function for this model was
 determined~\cite{bm}.
A key ingredient of this model was the   theory of
symmetric fields~\cite{mh,report}, which has been
used before in the spontaneous compactification and
dimensional reduction of higher dimensional models,
to obtain $SO(4)$-symmetric ans\"atze for the
fields in a $R\times S^3$ topology.
Let us remark here that it has been shown, at least in the case of
a scalar field coupled to gravity, that if one considers perturbations
of a minisuperspace model with three--spheres  as spatial sections,
any wave function, the object that we are mainly interested in,
has to be $SO(4)$ invariant~\cite{dowker}.

The main subject of this paper is precisely to consider
the above model beyond mini\-superspace, including all the modes of the fields.
For that purpose, we use the tool of harmonic analysis of symmetric fields on
homogeneous spaces~\cite{report,camp},
including to lowest non--trivial order the inhomogeneous
modes of the gauge fields, while treating the homogeneous degrees exactly.
The method of harmonic expansion around a symmetric
background configuration allows us to develop an analysis for
arbitrary gauge group. Furthermore, and more importantly,
it allows an analysis of the full realistic system
of gauge fields coupled to fermions and scalars in a unified way.
Therefore, using the method of symmetric fields one can consider the study of
the coupled matter sources as well as supergravity theories coupled to
Yang-Mills (YM), something which we consider as a continuation of our work.

The organization of our paper is the following. In section 2 we review the
construction of the symmetric model. In section 3 we apply the tools of the
harmonic analysis on homogeneous spaces and symmetric fields on our particular
system, i.e. a symmetric gauge field on $R\times S^3$ with
gauge group $G=SO(N)$.
Then we specialize on  the case $G=SO(3)$. In section 4 we
present the perturbed model and  in section 5 we consider the
equations of the tensor, vector and scalar modes and their
contribution to the wave function. We examine a
particular case, namely the full system of perturbations, when the classical
background field is fixed to its minimum value. Furthermore we
discuss the interactions between gauge and gravitational perturbations,
which might be quite interesting for the discussion of cosmological
models and anisotropies in the cosmological background radiation.
Finally in section 7 we present our conclusions.
We also provide two appendices with details about
the theory of symmetric fields and
harmonic analysis on $S^3$.

\section*{2.  General Formalism and the Symmetric Model }

Let us start by reviewing the relevant parts of the work already existing on
the subject; this will allow us to introduce the problem and fix
the notation.
We will deal with closed universes of the form $M=R \times B$,
where $B$ is a compact three-dimensional hypersurface which eventually
will be the three-sphere; time $t$ takes values in $R$.
For the description of such a universe, a $3+1$ formulation of the
four dimensional Einstein theory is appropriate. Such a description,
corresponding to an embedding of a three surface in the four dimensional
spacetime is given by the ADM formalism~\cite{adm}($\m, \n =0,\dots ,3,
\a,\b = 1,\dots,3$)~:
\beba
ds^2 &=& g_{\mu \nu}dx^\mu dx^\nu \\
&=& -(N^2 - N^\a N_\a)dt^2 + 2N_\a dx^\a dt +h_{\a\b} dx^\a dx^\b\,.
\label{metric}
\eaee
The quantities $N$ and $N_\a$ are called lapse and shift functions,
respectively, while $h_{\a\b}$ is a three-dimensional metric corresponding
to $B$. The action of the Einstein--Yang-Mills (EYM) system that we are going
to consider has generally
the form~:
\bea
S &=& S_{gr}+S_{YM}\,,
\label{a} \\
S_{GR} &=& {m_P^2 \over 16\pi}\left [\int _M d^4x~ (-g)^{1\over 2}~
 (R-2\lambda )
+2\int _{\partial M} d^3x~ h^{1\over 2}~ K\right ]\,,
\label{gr1}\\
S_{YM} &=& {1\over 8e^2}\int _M d^4x (-g)^{1\over 2}~
{\rm Tr}(F_{\mu\nu} F^{\mu\nu})\ ,
\label{ym}
\eea
where $R$ is the scalar curvature corresponding to $g_{\m\n}$, $h \equiv
{\rm det}(h_{\a \b})$
and $K$ is the trace of the extrinsic curvature $K_{\a\b}$
of $B \equiv \partial M$.
It may be shown that eq.~(\ref{gr1}) may be brought to the form
\be
S_{GR} = {m_P^2 \over 16\pi}\int d^3x dt
N~ [~G^{\a\b\g\d}~K_{\a\b}~K_{\g\d} + \sqrt{h}~ (^3R-2\lambda)~]
\,,
\label{gr2}
\ee
where $^3R$ is the curvature of $B$ and $G^{\a\b\g\d}$ is the DeWitt metric
given by
\be
G^{\a\b\g\d}={1 \over 2} \sqrt{h} (h^{\a\g} h^{\b\d} +h^{\a\d} h^{\b\g}
-2 h^{\a\b} h^{\g\d}).
\ee
 The corresponding $3+1$ - decomposition of eq.~(\ref{ym}) is
\bea
 S_{YM}&=&\frac{1}{8e^2}\int d^3xdt\sqrt{h}N{\rm tr}\left [\left (
h^{\a\g}h^{\b\d}
         -2h^{\a\g}h^{\b\e}h^{\d\r}\frac{N_\e N_\r}{N^2}\right
         )F_{\a\b}F_{\g\d} \right .\nnu \\
       &&\left .
-2\frac{h^{\a\b}}{N^2}F_{0\a}F_{0\b}-4h^{\a\b}h^{\g\d}\frac{N_\d}{N^2}
         F_{\a 0}F_{\b\g} \right ] \, . \label{ym2}
\eea

One may proceed as usual to determine the conjugate momenta and the
Hamiltonian.
It turns out that the Hamiltonian is a sum of constraints~:
\be
H=\int_{\partial M} d^3x \sqrt{h} (N~H_0~+~ N^\a~H_\a~+~{\rm tr}(A_0~H_{YM}))
\; .  \label{ham}
\ee
The lapse and shift functions play the role of Lagrange multipliers,
enforcing the constraints which correspond to Einstein's equations,
namely the momentum constraint $H_\a$ and the Hamiltonian constraint $H_0$.
Gauge invariance is guaranteed by the constraint $H_{YM}$ with the time
component $A_0$ of the gauge field as the corresponding Lagrange multiplier.

Now we are ready to consider the quantized model, which will be described
by a wave functional $\Psi[h_{\a\b}(x), A(x)]$
of the metric $h_{\a\b}(x)$ and the gauge field $A(x)$.
The constraints $H_0$, $H_\a$ and $H_{YM}$ are expressed by
demanding that their operator versions annihilate the wave functional
$\Psi[h_{\a\b}(x), A(x)]$.
They  ensure the invariance of the wave functional under
reparametrizations and gauge transformations.
The solution of the Wheeler-DeWitt equation $H_0\Psi = 0$ subject to the other
constraints lies at the heart of the subject of quantum
cosmology.

We proceed by reducing the infinite dimensional superspace $(h_{\a\b}(x),
A(x))$ to a finite dimensional set, the so called minisuperspace.
Some of the very common defining conditions of minisuperspace are~:
\begin{enumerate}
\item The lapse is taken to be homogeneous, i.e. $N=N(t)$.
\item The shift $N_\a$ is set to zero.
\item The three-metric $h_{\a\b}$ is constrained to be homogeneous and
isotropic.
\end{enumerate}
In quantum cosmology one is usually interested in closed universes. This,
together with the above requirements fixes the spatial sections of $M$ to be
$S^3$. Another way to single out this case is to impose $SO(4)\equiv
SU(2)\times SU(2)$
invariance on the metric of $B \equiv \partial M$ which leads to\footnote{
We will use Latin letters from the beginning of the alphabet to denote
representation indices taking values in the orthonormal bundle.}
\be
g_{00}=-\sigma^2 N(t)^2,~~
g_{ab} \equiv h_{ab} =\sigma^2 a(t)^2 \delta_{ab}, ~~
g_{0a} = 0\,.
\ee
An additional factor $\sigma ^2\equiv {8\pi / {m_P^2}}$ has been
introduced for convenience. Using this metric we obtain
\be
S_{GR_0} = -3 V_{S^3}\int  dt
\left [{a \over N} \dt a ^2 -N a +H^2 N a^3 \right ],
\ee
where $H^2 \equiv 8\pi\lambda/3 m_P^2$  and $V_{S^3}=2\pi^2$
is the volume of the three sphere with radius one.

For the matter sector, the following requirement is imposed, taking advantage
of the gauge freedom~: the change of the Yang-Mills field for motion on $B$
at fixed $t$ is allowed to be a gauge transformation rather than zero. In other
words, we are imposing $SO(4)\equiv SU(2)\times SU(2)$ invariance on the
gauge field modulo gauge transformation~\cite{report}.
For the gauge group $G=SO(N)$, according to the rules given in appendix A,
the gauge potential is
\be
A(t) \equiv A_0(t) dt + A_a(t) \omega^a,
\ee
with
\be
A_0(t)={1 \over 2} \Lambda^{IJ}(t)T_{IJ},
\label{a00}
\ee
\be
A_a(t)=[1+\c_0(t)]T_a + \c^I (t) T_{aI} \ ,
\label{aa0}
\ee
and the Maurer-Cartan forms $\o^a$ on the coset $S^3\cong SU(2)$ satisfying
$d\o^a+{\e^a}_{bc}\o^b\wedge\o^c =0$.
The matrices $T_{AB}$ are representation matrices of $SO(N)$
in correspondence with the decomposition  $SO(N) \supset SO(3)\times SO(N-3)$
\be\ba{ccccccc}
\ds {N(N-1)\over 2}&=& (3,1)&+&(1,\ds {(N-3)(N-4)\over 2})&+&(3,N-3) \ ,\\
                   &&T_a    &&   T_{IJ}                   &&T_{aI}
\label{dec}
\ea\ee
 which satisfy the following non-trivial commutation relations~:
\beba
\left [T_a, T_b \right]&=&\e _{abc} T_c \ ,\\
\left [T_a, T_{bK}\right ]&=&\e _{abc} T_{cK}\ ,\\
\left [T_{IJ}, T_{KL}\right ]&=&
-\d _{IK}T_{JL} +\d _{IL}T_{JK} +\d _{JK}T_{IL}
 -\d _{JL}T_{IK}\ ,\\
\left [T_{IJ}, T_{aK}\right ]&=& \d _{JK}T_{aI} -\d _{IK}T_{aJ}\,.
\label{son}
\eaee
 The functions
$\c_0(t), \c_I(t)$ and $\Lambda^{km}(t)$ are arbitrary.
We should also note that the isomorphic image of the
isotropy group $SO(3)$ in the gauge group $SO(N)$ is taken to be the group
generated by the three antisymmetric matrices $T_i$.
We see in this example how the $SO(4)$ symmetry requirement for the gauge
fields has reduced an infinite number of degrees of freedom down to a fairly
small number of functions. Moreover, for $SO(4)$ symmetric fields the
Lagrangian does not depend on the spatial coordinates, so the integration
over the three sphere will just yield the volume $V_{S^3}=2 \pi^2$
of $S^3$ as an overall factor.
 Taking this into account the action (\ref{ym2}),
after substituting
\bea
 F_{0a}&=&\dot{\c}_0 T_a+2\dot{\c}^I T_{aI}-2\c_I\Lambda^{IJ}T_{aJ}\ ,
 \label{F0_symm}\\
 F_{ab}&=&(1-\c_0^2-\chib^2)\e_{acb}T_c-2\c_0\e_{acb}\c^I T_{cI}
 \label{F_symm}
\eea
 for the field strength becomes
\be
S_{YM_0} = V_{S^3} { 3 \over {2 e^2} }\int dt \left[{a \over N}
\left( ({ {d \c_0} \over {dt} })^2 + (\cd_t
\chib)^2\right)
+2 {N \over a} V(\c_0, \chib)\right],
\label{ym_action0}
\ee
where
\be
\cd_t \chib \equiv {d \over {dt}} \chib + \Lambda \chib,
\ee
\be
V(\c_0,\chib) \equiv \frac{1}{2}(\c_0^2+\chib^2- 1)^2 +2\c_0^2\chib^2\; .
\label{ym_pot0}
\ee
Given the action $S_0=S_{GR_0}+S_{YM_0}$, we may find the conjugate momenta
and the Hamiltonian. It turns out that (dropping the factor $3V_{S^3}/e^2$
for simplicity)
\be
\pi_a=-{a \over N} \dt a, ~~\pi_{\c_0}={a \over N}  \dt {\c_0},~~
\pi_\chib = {a \over N} \cd_t \chib \,.
\ee
The Hamiltonian reads
\be
H \equiv {1 \over 2}{N \over a} \left[-\pi_a^2 - a^2 +H^2 a^4 +\pi_{\c_0}^2
+\pi_\chib^2 + 2V(\c_0,\chib)\right],
\ee
giving rise to the Wheeler-DeWitt equation
\be
{1 \over 2} \left[a^{-p} {\partial \over {\partial a}}
\left(a^p {\partial \over{\partial a}}\right)
-a^2 +H^2 a^4 -{{\partial^2} \over {\partial \c_0^2}}
-{{\partial^2} \over {\partial \chib^2}}+2V (\c_0, \chib)\right]
\psi(a, \c_0, \chib)=E \psi(a, \c_0, \chib)\; .
\ee
The parameter $p$ enters to take care of the operator ordering problem
\cite{hah}.
For the case under investigation the gauge constraints $H_{YM}$ are related to
the Lagrange multipliers $\Lambda$ (cf.~eq.~(\ref{ham})). They boil down to the
classical condition
\be
\c_I \pi_{\c_J} - \c_J \pi_{\c_I} = 0,
\ee
or, in the quantum language
\be
(\c_I {\partial \over {\partial \c_J}}
-\c_J {\partial \over {\partial \c_I}})\psi(a, \c_0, \chib)=0.
\ee
Based on the fact that the gravity and the gauge parts of the problem
decouple, the classical behavior has been determined for a general
$SO(N)$ gauge group. It exhibits interesting features, for example wormhole
solutions arise upon consideration of a Euclideanized version.
 Moreover, in the special case $N=3$, the wave function
has been determined in the semi--classical approximation by Bertolami and
Mour\~ao~\cite{bm}. Writing the separation ansatz
\be
\Psi(a,\c_0)=\sum_n^{\,}C_n(a)U_n(\c_0)\ ,
\label{decoup}
\ee
where $C_n(a)$ is the gravitational wave--function obtained by Hartle
and Hawking~\cite{hh}, $U_n(\c_0)$ is given by
\beba
U(\c_0) &=& A \exp \left( - \frac{V_{S^3}}{e^2}(3(\c_0 +1)^2
            - (\c_0 +1)^3) \right)\,,
\quad{\rm for}\quad \c_0 <-1\ ,\\
 U(\c_0) &=& B \exp \left( - \frac{V_{S^3}}{e^2}(3(\c_0 +1)^2
             - (\c_0 +1)^3)\right) \\
 &\quad+& B \exp \left( - \frac{V_{S^3}}{e^2}(3(\c_0 -1)^2
         + (\c_0 -1)^3)\right)\,,
 \quad{\rm for}\quad|\c_0| \leq 1\ ,\\
 U(\c_0) &=& A \exp \left( - \frac{V_{S^3}}{e^2}(3(\c_0 -1)^2
             + (\c_0 -1)^3)\right)\,,
 \quad{\rm for}\quad \c_0 >1\; .
\label{solut}
\eaee
Here $A=(1+\exp(-4/3))B$ and $B$ is a normalization factor. The Hartle--Hawking
boundary proposal has been imposed to arrive at these solutions by
requiring that $\c_0(\eta=-\infty) = \mp 1$.
Furthermore $\c_0(\eta=0)=\c_0$ is the value at the given three--surface
and $\eta$ is the conformal
time.

\section*{3. Harmonic Analysis}
\noindent
A very interesting issue is to see what happens beyond the
minisuperspace approximation. This has been done by Halliwell and Hawking
\cite{hh} for the case of a scalar field coupled to gravity.
They consider perturbations around the minisuperspace model, considering
the following forms for the various fields involved
\be
h_{\a\b}= a^2 (\Omega_{\a\b}+\ep_{\a\b}), ~~\phi(y,t)=\phi(t)
+\delta \phi (y,t),~~
N(y,t)=N_0(t)+\delta N(y,t),
\ee
where $\Omega_{\a\b}$ is the metric of the three sphere,
allowing also $N_\a(y,t)$ to take (small) non-zero values~\footnote{Here and
in the following $y$ denotes the coordinates of the three sphere.}. The
background quantities $ a,\phi,N_0 $ are treated to all orders, and the
perturbations up to second order. The method to handle the perturbations
is the use of harmonic expansion.
In this paper we would like to examine the problem of perturbations
around minisuperspace for the case of the $SO(4)$ symmetric gauge fields
referred to above. We therefore split the gauge field
\be
 A = A^{(0)} + \bar{A}
\ee
into a symmetric part $A^{(0)}$ as it was presented for the group $G=SO(N)$
in
the last section, and a perturbation $\bar{A}$. On the other hand, following
appendix A, $A_0$ (which behaves as a scalar as far as the $SO(4)$ rotations
are concerned) and $A_a$ can be written in the form
\begin{eqnarray}
A_0 (t,y) &=&  \sum _{m,pq} ~D_{pq}^{(m)}(L(y))~ a_0 ~^{(m)}_{pq}(t)
\,,
\label{aoexp}\\
A_a(t,y) &=& A_a^{\sis B}+\sum _{m,pq}
 ~D_{pq}^{(m)}(L(y))~ \c _a ~^{(m)}_{pq}(t)\, .
\label{anal}
\end{eqnarray}
We note that the factors $\sqrt{ {{d_m} \over {d_{\cd}}} }$ appearing in
equations (\ref{349}) and (\ref{3410}) have been absorbed
in the coefficients $a_0 ~^{(m)}_{pq}(t)$ and $\c _a ~^{(m)}_{pq}(t).$
In these equations the field $A^{(0)}$ can be identified with the background
$A^B$ plus the $y$-independent parts of the expansions corresponding to the
trivial representation $(m=0)$ of $SO(4)$.
In order to specify the harmonic analysis of the gauge field we must first
solve the constraints given in appendix A for the $D_{pq}^{(m)}$ $SO(4)$
representation matrices.

We decompose the adjoint of the gauge group $G$ under $R=SO(3)
\approx SU(2)$ according to the embedding~:
\beba
G &\supset & R \times H\ ,\\
{\rm adj}~G &=& (3,1)+(1,{\rm adj}~H)+\sum_k~(R_k,H_k)\; .
\label{decg}
\eaee
Let us next introduce the generators $T_a \,, T_L$ and $T_{r_k h_k}$
of $G$ in correspondence with the decomposition (\ref{decg}).  We write
their nontrivial commutation relations as
\bea
 \left [T_a,T_b \right ]&=&\e_{abc}T_c \ ,\\
 \left [T_a,T_{r_k h_k}\right ]&=& D_{r'_k r_k}^{(R_k)}(T_a)T_{r'_k h_k}\, .
\eea
 The coefficients $a_0 ~^{(m)}_{pq}(t)$ and $~\c_a~^{(m)}_{pq}(t)$
take values in the Lie algebra of $G$ and
may be expanded in the basis
of the matrices $T_a, T_L$ and $T_{r_k h_k}$ defined above.
The relevant expansions read~:
\bea
a_0 ~^{(m)}_{pq}(t)~&=~&\sum_c \alpha_{pq|c}^{(m)}(t) T_c
                   +\sum_L \alpha_{pq|L}^{(m)}(t) T_L
+\sum_{k , r_k , h_k}\alpha _{pq|r_k h_k}^{(m)(R_k,H_k)}(t) T_{r_k h_k}
\label{aaa0}\ ,\\
\c_a ~^{(m)}_{pq}(t)~&=~&\sum_c \c_{a,pq|c}^{(m)}(t) T_c
                   + \sum_L \c_{a,pq|L}^{(m)}(t) T_L
+\sum_{k , r_k , h_k} \c_{a,pq|r_k h_k}^{(m)(R_k,H_k)}(t) T_{r_k h_k}\; .
\label{chia}
\eea
In (\ref{chia}) the indices $r_k, h_k$ run over the representations
$R_k, H_k$ respectively.

The above expansion has been made with an eye to the solution of the
constraints (\ref{3414}) and (\ref{3415})
 in appendix A. Indeed, the constraints can be
expressed in terms of the components of $a_0 ~^{(m)}_{pq}(t)$ and
$\c_a ~^{(m)}_{pq}(t)$ as follows (summation over repeated indices is
implied)~:
\beba
D_{ps}^{(m)}(T_c) \alpha_{pq|a}^{(m)}(t)-\ep_{cba}\alpha_{sq|b}^{(m)}(t)
&=&0\ ,\\
D_{ps}^{(m)}(T_c)\alpha_{pq|J}^{(m)}(t)&=&0\ ,\\
D_{ps}^{(m)}(T_c)~\alpha_{pq|r_k h_k}^{(m)(R_k,H_k)}(t)
-D_{r_k^\prime r_k}^{(R_k)}(T_c)
\alpha_{sq|r_k^\prime h_k}^{(m)(R_k,H_k)}(t) & = & 0\ ,
\eaee
\beba
D_{ps}^{(m)}(T_c)\c_{a,pq|d}^{(m)}(t)
+ \ep_{cab} \c_{b,sq|d}^{(m)}(t)
+ \ep_{cdb} \c_{a,sq|b}^{(m)}(t)&=&0\ ,\\
D_{ps}^{(m)}(T_c)\c_{a,pq|J}^{(m)}(t)
+ \ep_{cab} \c_{b,sq|J}^{(m)}(t)&=&0\ ,\\
D_{ps}^{(m)}(T_c)~\c_{a,pq|r_k h_k}^{(m)(R_k,H_k)}(t)
+\ep_{cab}~\c_{b,sq|r_k h_k}^{(m)(R_k,H_k)}(t)& & \\
-D_{r_k^\prime h_k}^{(R_k)}(T_c)~\c_{a,sq|r_k^\prime h_k}^{(m)(R_k,H_k)}(t)
&=&0 \; .
\eaee
In this form it is easy to invoke Schur's lemma and find the solution of
the constraints. According to the discussion in appendix B, we denote
the representation $(m)$ by $(j_L,j_R|J)$; in this notation the nontrivial
solution is written as
\beba
\a_{sq|c}^{(j_L,j_R|1)}&=&\delta_{sc} \alpha_q^{(j_L,j_R|1)}\ ,\\
\a_{0q|L}^{(j_L,j_R|0)}&=&\a_{q|L}^{(j_L,j_R|0)}\ ,\\
\a_{sq|r_k h_k}^{(j_L,j_R|J)(R_k,H_k)}&=&\delta_{s r_k}
\a_{q|h_k}^{(j_L,j_R|J)(R_k,H_k)}, \quad {\rm if} \quad
(j_L,j_R) \supset R_k\ ,
\eaee
\beba
\c_{a,sq|b}^{(j_L,j_R|J)}&=&<1a,1b|Js> \c_q^{(j_L,j_R|J)}, J=0,1,2\ ,\\
\c_{a,sq|L}^{(j_L,j_R|1)}&=&\delta_{as} \c_{q|L}^{(j_L,j_R|1)}\ ,\\
\c_{a,sq|r_k h_k}^{(j_L,j_R|J)(R_k,H_k)}
&=&<1a,R_k r_k|Js> \c_{q|h_k}^{(j_L,j_R|J)(R_k,H_k)}\; .
\label{crkhk}
\eaee
In the last equation $J$ takes the values $j_k-1, j_k, j_k+1$, where $j_k$
is the angular momentum describing the representation $R_k$.  We conclude
that the possible values of $J$ which determine the degree of harmonics which
appear in the harmonic expansion are $\{ 0,1,2,j_k-1,j_k,j_k+1\}$.
In particular:
\be
J_{max} = {\rm max}(2,j_{max}+1)\ ,
\ee
where $j_{max}$ is the maximum angular momentum corresponding to the
representations $R_k$ which appears in the decomposition (\ref{decg}) of $G$
under $SO(3)$. The harmonic expansions read~:
\beba
A_0(t,y)&=&\sum_{(j_L,j_R) \supset 1} \sum_{b,q}
D_{bq}^{(j_L,j_R|1)}(L(y)) \a_q^{(j_L,j_R|1)}(t) T_b\\
 &+&\sum_{(j_L,j_R) \supset 0} \sum_{q,L}
D_{0q}^{(j_L,j_R|0)}(L(y)) \a_{q|L}^{(j_L,j_R|0)}(t) T_L\\
 &+&\sum_{(R_k,H_k)} \sum_{q,(j_L,j_R) \supset R_k}
D_{r_k q}^{(j_L,j_R|R_k)}(L(y)) \a_{q|h_k}^{(j_L,j_R|R_k)(R_k,H_k)}(t)
T_{r_k h_k}\ ,
\label{a0ty}
\eaee
\beba
A_a(t,y) & = & A_a^B\\
 &+&\sum_{J=0,1,2} \sum_{(j_L,j_R) \supset J} \sum_{b,p,q}
D_{pq}^{(j_L,j_R|J)}(L(y)) <1a,1b|Jp> \c_q^{(j_L,j_R|J)}(t) T_b\\
 &+&\sum_{(j_L,j_R) \supset 1} \sum_{q,L}
D_{aq}^{(j_L,j_R|1)}(L(y)) \c_{q|L}^{(j_L,j_R|1)}(t) T_L\\
 &+&\sum_{(R_k,H_k)} \sum_{J=R_k-1,R_k,R_k+1}\sum_{(j_L,j_R) \supset J}
 \sum_{p,q,r_k,h_k} D_{pq}^{(j_L,j_R|J)}(L(y))\\
 & & <1a,R_k r_k|Jp>
\c_{q|h_k}^{(j_L,j_R|J)(R_k H_k)}(t) T_{r_k h_k} \; .
\eaee
The expression $D_{0q}^{(j_L,j_R|0)}(L(y))$ on the second line of
(\ref{a0ty}) just means that the first index of this harmonic takes the
value corresponding to the identity representation in the product
$(j_L, j_R)$. On the other hand the notation
$\sum_{(j_L,j_R) \supset J}$ means that we sum over all the tensor
product representations $(j_L, j_R)$ of $SO(4) \approx
SU(2)_L \otimes SU(2)_R$, which contain the representation $J$ upon
restriction to $SU(2)_{diag}$, as described in appendix B.

If the gauge group $G$ is $SO(N)$, several simplifications occur.
The decomposition of the adjoint of $SO(N)$ under $R=SO(3)$ can be written
in a more concrete form and it is given in eq.~(\ref{dec}).
The commutation relations of the generators, heavily used in the calculations,
corresponding to the decomposition (\ref{dec}) can be found in
eqs.~(\ref{son}).

Maybe the most important simplification brought about is that the only
representation appearing in the decomposition (\ref{dec}) is
the triplet, corresponding to angular momentum $j_{max}=1$.
This makes things much easier, since according to the
statements made above only scalar- vector- and tensor-harmonics
appear in the expansion. In this case we are also in a position to
give explicit expressions for the relevant Clebsch-Gordan coefficients.
 Changing from the $SU(2)$ standard basis to the ``tensor'' basis
they read~:
\beba
<1a,1b|00>~&\rightarrow~& \delta_{ab}\\
<1a,1b|1p>~&\rightarrow~& \ep_{abp}\\
<1a,1b|2~p_1 p_2>~&\rightarrow~&{1 \over 2} (\delta_{a p_1} \delta_{b p_2}+
\delta_{b p_1} \delta_{a p_2})
-{1 \over 3} \delta_{ab} \delta_{p_1 p_2}\; .
\label{cgcoeff}
\eaee
All indices run from 1 to 3 and we have represented the $J=2$ components by
the double index $p_1 p_2$, as in Appendix B.

  We are going to discuss in the following section
the case $N=3$. Most of the physical features of the model are present in
this case; on the other hand, calculations are much easier,
because of some additional simplifications. Also the formulae to be
presented, although far from being simple, are much more economical.
The simplifications coming from choosing $N=3$ are summarized below.

One may see that the parts of the expansion
having to do with ${\rm adj}~SO(N-3)$ and $(R_k,H_k)$ (cf.~eq.~(\ref{decg}))
no longer exist. Thus the harmonic expansion becomes~:
\beba
A_0(t,y)&=&\sum_{(j_L,j_R) \supset 1} \sum_{b,q}
D_{bq}^{(j_L,j_R|1)}(L(y)) \a_q^{(j_L,j_R|1)}(t) T_b\ ,
\eaee
\beba
A_a(t,y) & = & A_a^B
 +\sum_{J=0,1,2} \sum_{(j_L,j_R) \supset J} \sum_{b,p,q}
D_{pq}^{(j_L,j_R|J)}(L(y))\\
 & & <1a,1b|Jp> \c_q^{(j_L,j_R|J)}(t) T_b \; .
\eaee
We separate the background field and the $y$-independent part
in the last equation and write explicitly the sum over $J$~:
\beba
A_a(t,y) & = & (1 + \c_0(t)) T_a\\
 &+&\sum_{(j_L,j_R) \supset 0} \sum_{b,q}
D_{0q}^{(j_L,j_R|0)}(L(y)) \delta_{ab} \c_q^{(j_L,j_R|0)}(t) T_b\\
 &+&\sum_{(j_L,j_R) \supset 1} \sum_{b,p,q}
D_{pq}^{(j_L,j_R|1)}(L(y)) \ep_{abp} \c_q^{(j_L,j_R|1)}(t) T_b\\
 &+&\sum_{(j_L,j_R) \supset 2} \sum_{b,q,p_1,p_2}
[{1 \over 2} (\delta_{a p_1} \delta_{b p_2}
 + \delta_{b p_1} \delta_{a p_2})\\
 &-&{1 \over 3} \delta_{ab} \delta_{p_1 p_2}]
D_{(p_1 p_2)q}^{(j_L,j_R|2)}(L(y)) \c_q^{(j_L,j_R|2)}(t) T_b,
\eaee
where we have plugged in the explicit expressions (\ref{cgcoeff}) for the
Clebsch-Gordan coefficients. The Clebsch-Gordan coefficient
${1 \over 2} (\delta_{a p_1} \delta_{b p_2}+
\delta_{b p_1} \delta_{a p_2})
-{1 \over 3} \delta_{a b} \delta_{p_1 p_2}$
in fact coincides with the operator $P(p_1 p_2|q_1 q_2)$, which projects
out the part of the harmonic
$D_{(p_1 p_2)q}^{(j_L,j_R|2)}(L(y))$ which is symmetric and
traceless in the indices $p_1$ and $p_2$. Thus, this Clebsch-Gordan
coefficient just projects
out the harmonics $G_{ab|q}^{(n \pm)}(L(y))$, $S_{ab|q}^{(n \pm)}(L(y))$ and
$P_{ab|q}^{(n)}(L(y))$, discussed in Appendix B, where it is also
explained that the harmonics $D_{aq}^{(j_L,j_R|1)}(L(y))$ will give rise to
$S_{a|q}^{(n \pm)}(L(y))$ and $P_{a|q}^{(n)}(L(y))$. Similarly, the harmonics
$D_{0q}^{(j_L,j_R|0)}(L(y))$ become $Q^{(n)}_q(L(y))$.
Then, introducing some numerical factors for later convenience,
the final form of the harmonic expansion in this case reads
\beba
A_0(t,y)&=&\sum_n \sum_{b,q}
[\sqrt{2}\a^{(n+)}_q(t)S_{b|q}^{(n+)}(L(y))+
\sqrt{2}\a^{(n-)}_q(t)S_{b|q}^{(n-)}(L(y))\\
 &+&{1\over\sqrt{6}}\b^{(n)}_q (t)P_{b|q}^{(n)}(L(y))] T_b
 \label{hara0}
\eaee
\beba
A_a(t,y) & = &  A_a^{(0)}\\
 &+&\sum_{n} \sum_{b,q}
{1\over\sqrt{6}}\gamma^{(n)}_q(t) \delta_{ab} Q^{(n)}_q (L(y)) T_b\\
 &+&\left [ \sum_{n} \sum_{b,q}
{1\over\sqrt{2}}\rho^{(n+)}_q(t)S_{c|q}^{(n+)}(L(y))
+{1\over\sqrt{2}}\rho^{(n-)}_q(t)S_{c|q}^{(n-)}(L(y))\right .\\
 &+&\left . \sum_{n} \sum_{b,q}
{1\over\sqrt{6}}\sigma^{(n)}_q(t)P_{c|q}^{(n)}(L(y))\right ] \ep_{acb} T_b\\
 &+&\sum_{n} \sum_{b,q}
[\mu^{(n+)}_q(t)G_{ab|q}^{(n+)}(L(y))+\mu^{(n-)}_q(t)G_{ab|q}^{(n-)}(L(y))]
 T_b\\
 &+&\sum_{n} \sum_{b,q}
{1\over\sqrt{2}}[\nu^{(n+)}_q(t)S_{ab|q}^{(n+)}(L(y))
 +\nu^{(n-)}_q(t)S_{ab|q}^{(n-)}(L(y))]T_b\\
 &+&\sum_{n} \sum_{b,q}
\sqrt{6}[\xi^{(n)}_q(t)P_{ab|q}^{(n)}(L(y))] T_b\ ,
\label{haraa}
\eaee
 with
\be
 A_a^{(0)} =  (1 + \c_0(t)) T_a\, .
\ee

\section*{4. The Perturbed Model}

Let us now consider inhomogeneous perturbations around our minisuperspace.
We use the exact action in terms of the
background quantities but expand only to second order in the perturbations.
We assume that the metric is of the form given in eq.~(\ref{metric}) except
that it is multiplied with the normalization factor $\sigma ^2$.
The gravitational part of our model is treated in exactly the same way
as in ref.~\cite{hh}. Therefore let us recall that
\be
h_{ab} = \s^2 e^{2\alpha}~(\d _{ab} + \ep _{ab})\,,
\label{h}
\ee
where $e^\alpha =a$ and (in simplified notation)
\be
\ep _{ab}= {\sqrt 6\over 3}~\d _{ab}~ a_n Q^n +
 {\sqrt 6}~ b_n P_{ab}^n+{\sqrt 2}~ c_n S_{ab}^n+
2 ~d_n G_{ab}^n\,.
\label{eh}
\ee
Furthermore for the lapse and shift functions we have~:
\bea
N &=& N_0~ \left [ 1+{1\over \sqrt 6}~g_n Q^n\right ]\,,
\label{no}\\
N_a &=& e^\alpha ~\left [{1\over \sqrt 6} ~k_n P_a^n +{\sqrt 2}~j_n
S_a^n\right ]\,.
\label{ni}
\eea
Let us remark here that, of course, these expansions are exactly
the same as the ones we get, if we apply the symmetric field constraints
on the functions $\ep _{ab}$, $N$ and $N_a$.
Now the gravitational part of the action (\ref{gr2}) takes the form~\cite{hh}
\be
S_{GR}=S_{GR_0}+V_{S^3}\sum _n \int dt L_{GR}^n \, ,
\ee
where
\begin{equation}\begin{array}{lccl}
L_{GR}^n &=&
\ds {{1\over 2}} e^\alpha N_0 &\left[  \ds{
{1\over 3}(n^2 -{5\over 2})a_n^2+{n^2-7\over
3}~ \frac{n^2-4}{n^2-1}b_n^2
-2~(n^2-4)~c_n^2 }\right. \\
&&&\quad \ds{ -(n^2+1)d_n^2 +{2\over
3}(n^2-4)a_nb_n +{2\over 3}(n^2-4)g_nb_n}\\
&&&\quad \left. \ds {+ {2\over 3}(n^2
+{1\over2})g_n a_n} \right ]\\
&&\ds{ +
2 {e^\alpha \over N_0}}& \left [ \ds{
- {1\over 3(n^2-1)}k_n^2 +(n^2-4)j_n^2}\right ] \\
&& \ds{+
{1\over 2} {e^{3\alpha} \over  N_0}}& \left \{ \ds{
-\dt a_n^2 + \frac{ n^2-4}{ n^2-1}\dt b_n^2 +
 (n^2-4) \dt c_n^2 +\dt d_n^2 }\right.\\
&&&\quad +\dt
\alpha \left [  \ds{ -2a_n \dt a_n +8 {n^2-4\over n^2-1}
b_n \dt b_n +8 (n^2-4)c_n \dt c_n +8d_n \dt d_n +2g_n \dt a_n}\right ]\\
&&&\quad +\dt \alpha^2 \left [ \ds{ -{3\over 2}a_n^2 +
6{ n^2-4\over
n^2-1}b_n^2+6(n^2-4)c_n^2+6d_n^2 +3a_n g_n -g_n^2}\right ]\\
&&& \quad
+e^{-\alpha}\left.
\left [ \ds {k_n \left ( -{2\over 3} \dt a_n -{2\over 3} {
 n^2-4\over n^2-1} \dt b_n +{2\over 3} \dt \alpha g_n\right )
-2(n^2-4)\dt c_n j_n}\right ]\right \}\\
&& \ds{- 3 e^{3\alpha} N_0 }& H^2 \left[ \ds{
{1\over 4} a_n^2 -{n^2-4 \over n^2-1} b_n^2 -(n^2-4) c_n^2 -d_n^2
 +{1\over 2}a_n g_n}\right]\, .
\label{lgr}
\end{array}\end{equation}

 For the YM part of the action with gauge group $G=SO(3)$
we write the field strength as
\beba
 F_{a0}&=&F_{a0}^{(0)}+\nabla_a^{(0)}\bar{A}_0-\nabla_0^{(0)}\bar{A}_a
          +[\bar{A}_a,\bar{A}_0]\ ,\\
 F_{ab}&=&F_{ab}^{(0)}+\nabla_a^{(0)}\bar{A}_b-\nabla_b^{(0)}\bar{A}_a
          +[\bar{A}_a,\bar{A}_b]\ ,
\label{F}
\eaee
with the symmetric field $F^{(0)}$ given in eq.~(\ref{F_symm}) and a
covariant derivative with respect to the symmetric field~:
\be
 \nabla_a^{(0)} = \nabla_a+[A_a^{(0)},\cdot]\ , \quad\quad
 \nabla_0^{(0)} = \partial_0\, .
\ee
Inserting the symmetric field (\ref{aa0}) and the harmonic expansions
(\ref{hara0}),(\ref{haraa}) together with the above expressions in
eq.~(\ref{ym2}) we obtain\footnote{Here we give the Minkowskian action.
For the Euclidean version the signs of ${\cal L}_2^n$ and ${\cal L}_3^n$
have to be reversed.}
\be
 S_{YM}=S_{YM_0}+{V_{S^3}\over e^2}\sum _n \int dt L^n_{YM}\ ,
\ee
with
\begin{equation}
 L_{YM}^n =  \ds{N_0\over e^{\alpha}} {\cal L}_1^n
+  \ds{{e^{\alpha}\over N_0}} {\cal L}_2^n
+ \ds{{1\over N_0{e^{\alpha}}}}{\cal L}_3^n
+ \ds {1\over N_0}  {\cal L}_4^n
\label{lagr}
\end{equation}
and
\begin{equation}\begin{array}{lcl}
{\cal L}_1 &=&  \ds{ (\chi_0 ^2 -1)^2 \left(
-{3\over 8} a_n^2 +{1\over
 4} a_ng_n -{3\over 2} {n^2-4 \over n^2-1}b_n^2 - {3\over 2} d_n^2 -
 {3\over 2} (n^2-4)c_n^2 \right)} \\
&&\ds {+\chi_0  (\chi_0^2-1)\left(
a_n \gamma_n +4{n^2-4 \over n^2-1} b_n \xi_n
+2(n^2-4)c_n \nu_n+2d_n \mu_n -g_n\gamma_n\right)}\\
&&\ds {+\chi_0^2\left(
-{3\over 2}\gamma_n^2 -\rho_n^2 -{1\over 3}{1\over n^2-1}\sigma _n^2
+{1\over 3}a_n\sigma_n -{2\over 3}  {n^2-4 \over n^2-1} b_n \sigma_n\right.}\\
&&\ds {\quad \quad \quad\left.
-(n^2-4)c_n \rho_n -{1\over 3}g_n \sigma _n +n(n^2-4)\td c_n \nu_n
+2n \td d_n \mu_n\right)} \\
&& \ds {+\chi_0\left(
-\gamma_n\sigma _n + {1\over 2}n \td \rho_n \rho _n -n \td \mu_n \mu_n
-{1\over 2} n(n^2-4)\td\nu_n\nu_n\right)} \\
&&\quad \ds
-{1\over 6} (n^2-4) \gamma_n^2 - {1\over 4} (n^2-4) \rho_n^2
-{1\over 6}  {n^2-3 \over n^2-1} \sigma_n^2
-{1\over 4} (n^2-4)n^2\nu_n^2\\
&&\quad \ds
-{1\over 2} (n^2+1)\mu_n^2
-{2\over 3} (n^2-4)\xi_n^2
-{1\over 3}  a_n \sigma_n
+{2\over 3} {n^2-4 \over n^2-1} b_n\sigma_n
+(n^2-4)c_n\rho_n  \\
&&\quad \ds
+{1\over 3} g_n \sigma_n
-n(n^2-4)\td c_n \nu_n
-2n\td d_n \mu_n
-{2\over 3}(n^2-4) \gamma_n \xi_n
+{1\over 2} n(n^2-4)\td \rho_n \nu_n \ ,
\end{array}\end{equation}
\begin{equation}\begin{array}{lcl}
{\cal L}_2 &=&  \ds
\chi_0^2\left(  2 \alpha_n^2   + {1\over 6}{1\over n^2-1} \beta_n^2\right)
+\chi_0\left(-2 \alpha_n \dot{\rho_n}
-{1\over 3}{1\over n^2-1} \beta_n \dot{\sigma_n}
+2n\td \alpha _n \alpha _n \right)\\
&& \ds
+ \dot{\chi}_0^2 \left(
 -{1\over 8} a_n^2 + {1\over 2}  {n^2-4 \over n^2-1} b_n ^2
+ {1\over 2} (n^2-4) c_n^2 + {1\over 2} d_n^2 + {1\over 4} g_n^2
 - {1\over 4}a_n g_n \right)   \\
&& \ds
+ \dot{\chi}_0 \left(
 {1\over 6} a_n \beta_n   + {1\over 2} a_n \dot{\gamma_n}
+ {2\over 3}  {n^2-4 \over n^2-1}b_n\beta_n
- 4  {n^2-4 \over n^2-1} b_n\dot{\xi_n}
+ 2(n^2-4)  c_n \alpha_n \right.\\
&& \ds    \left.
- 2 (n^2-4) c_n \dot{\nu_n}
- 2 d_n\dot{\mu_n}  -  {1\over 6} g_n \beta_n
-  {1\over 2} g_n \dot{\gamma_n}   + 2  \alpha_n \rho_n
+  {1\over 3}  {1\over n^2-1}\beta_n \sigma_n \right)\\
&& \ds
+ (n^2-2)\alpha_n^2  - (n^2-4)\alpha_n \dot{\nu_n}
+  {1\over 12}  {n^2-3 \over n^2-1}  \beta_n^2
+  {1 \over 6} \beta_n \dot{\gamma_n}
-  {2 \over 3}{n^2-4\over n^2-1} \beta_n \dot{\xi_n}\\
&& \ds
+  {1 \over 4}\dot{\gamma_n}^2
+  {1 \over 2} \dot{\rho_n}^2
+ {1 \over 6}{1\over n^2-1} \dot{\sigma_n}^2
+  {1 \over 2} \dot{\mu_n}^2
+  {1 \over 2}(n^2-4) \dot{\nu_n}^2
+ 2{n^2-4\over n^2-1} \dot{\xi_n}^2
- n \td \alpha_n \dot{\rho_n} \ ,
\end{array}\end{equation}
\begin{equation}
{\cal L}_3 =\ds  (\chi_0^2-1)^2
\left( {1\over 6(n^2-1)} k_n^2 + 2 j_n^2 \right)\ ,
\end{equation}
\begin{equation}\begin{array}{lcl}
{\cal L}_4  &=&  \ds
\chi_0 \dot{\chi}_0 \left(-{1\over 3}{1\over n^2-1} k_n \sigma_n - 2 j_n
\rho_n\right)
+\chi_0 (\chi_0^2-1) \left(-{1\over 3}{1\over n^2-1} k_n \beta_n
- 4 j_n \alpha_n\right)
\\
&& \ds
+\dot{\chi}_0 \left(-{1\over 3} k_n \gamma_n -{2\over 3}{n^2-4 \over n^2-1}
  k_n \xi_n-(n^2-4)j_n \nu_n+n\td j_n \rho_n \right)
\\
&&\ds
+(\chi_0^2-1)\left({1\over 3}{1\over n^2-1} k_n \dot{\sigma_n}
+ 2 j_n \dot{\rho_n}  - 2 n \td j_n \alpha_n \right)\, .
\end{array}\end{equation}
 In these (and the following) formulae a sum over the even and the odd
part of the relevant coefficients is implied.
The tilde over a coefficient signals an
off-diagonal term connecting even and odd perturbations with each other
\footnote{This mixing of the odd and even parts may give rise to questions
about parity conservation when the perturbations are included. This does
not create any problem though:
If we start from the parity
transformation properties of the metric and the gauge fields, we may
derive the ones of $N_0(t)$, $a(t)$, $\chi_0(t)$ and the perturbation
coefficients. Using these properties, it turns out that the second-order
Lagrangian is parity invariant.}:
$\tilde{p}_n^{(even/odd)}=p_n^{(odd/even)}$ for a generic
coefficient $p_n$.

 One may ask about the remnants of gauge symmetry in this Lagrangian. To
answer this question we start with an ordinary infinitesimal gauge
transformation $\d A_\m = \nabla_\m U-[U,A_\m]$, $U=u_aT_a$ and demand that
the symmetric field $A^{(0)}$ remains fixed. The gauge transformation will
therefore exclusively act on the expansion $\bar{A}$ and we find
$\d A_\m=\nabla_\m^{(0)}U$ in the leading order. Expanding the parameter $u_i$
\be
 u_a = \frac{1}{\sqrt{6}}v_n P_a^n+\sqrt{2}w_nS_a^n
 \label{u_exp}
\ee
and using the expression (\ref{aa0}) for the symmetric field as well as
eqs.~(\ref{hara0}) and (\ref{haraa}) for $\bar{A}$ this transformation
property can be translated to the coefficients~:
\be
 \label{trafo}
 \ba{llllllllllll}
 \d\g_n&=&-\frac{1}{3}v_n \ ,
 &\d\x_n&=&\frac{1}{6}v_n \ ,
 &\d\s_n&=&\c_0v_n \ ,
 &\d\b_n&=&\dot{v}_n , \\
 \d\n_n&=&w_n \ ,
 &\d\a_n&=&\dot{w}_n \ \ ,
 &\d\r_n&=&2\c_0 w_n+n\tilde{w}_n \ ,
 &\d\m_n&=&0 \, .
\ea
\ee
Our procedure is analogous to the definition of coordinate transformations
acting on the perturbations of a certain background metric as given in
ref.~\cite{bardeen}. This analogy can be pushed one step further by
introducing gauge invariant variables
\be
 \label{inv}
 \ba{lllllllll}
 \Gamma_n&=&\g_n+2\x_n \ ,
 &B_n&=&\b_n+3\dot{\g}_n \ ,
 &S_n&=&\s_n+3\c_0\g_n \ , \\
 A_n&=&\a_n-\dot{\n}_n \,
 &R_n&=&\r_n-2\c_0\n_n-n\tilde{\n}_n \ . &&&
 \ea
\ee
As in the case of coordinate transformations in pure gravity the tensor
mode $\m$ is invariant, since the expansion~(\ref{u_exp})
does not contain any tensor degree of freedom. It might be interesting
to note~\footnote{We thank C.~Lee for pointing out this relation.}
that the above variables represent a linearized version of the invariant
quantities tr$(B_cB_d)$ introduced by L\"uscher~\cite{luscher}
($B_c=\e_{acb}F_{ab}$ is the magnetic field.). With these variables one
gets, for instance:
\be
 R_n=\frac{\sqrt{2}}{8(n^2-4)(\c_0^2-1)}\int d\m (y)
      {\rm tr}(B_cB_d)S_{cd}^n\, .
\ee

The important observation is now that our Lagrangian (\ref{lagr}) is
invariant under the transformations (\ref{trafo}) provided the field
$\c_0$ fulfills the background
equation of motion to be derived from the action (\ref{ym_action0}).
Consequently, in such a case the Lagrangian can be written in terms of
the gauge invariant variables only. For a special solution $\c_0$ we will
demonstrate this explicitly in section 5.

One can proceed from this point by defining the conjugate momenta and
obtaining the Hamiltonian in the usual manner.
Here we want to indicate the structure of the Hamiltonian
\be
 H = N_0(\ch_0+\sum_n \ch_2^{(n)}+\sum_n g_n \ch_g^{(n)})
 +\sum_n(k_n\ch_k^{(n)}+j_n\ch_j^{(n)})
 +\sum_n(\a_n\ch_\a^{(n)}+\b_n\ch_\b^{(n)})\ ,
 \label{ham_str}
\ee
which we find in accordance with eq.~(\ref{ham}). The linear YM gauge
constraints $\ch_\a^{(n)}$ and $\ch_\b^{(n)}$ responsible for the gauge
invariance of the wave function read
\beba
 \ch_\a^{(n)}&=&2\c_0\p_{\r_n}-{2 \over 3}\r_n\p_\c
                +\p_{\n_n}+n\tilde{\p}_{\r_n}\ ,\\
 \ch_\b^{(n)}&=&\c_0\p_{\s_n}-\frac{1}{9(n^2-1)}\s_n\p_\c-{1 \over 3}\p_{\g_n}
         +{1 \over 6}\p_{\x_n}\ ,
 \label{ham_gau}
\eaee
with the conjugate momenta
\beba
 \p_\c&=&\frac{3V_{S^3}e^\a}{e^2N_0}\dot{\c_0}+({\rm quadratic}\;{\rm terms})\
, \\
 \p_{\g_n}&=&\frac{V_{S^3}e^\a}{2e^2N_0}\left[\dot{\g}_n+{1 \over 3}\b_n
             +\dot{\c_0}(a_n-g_n)\right] \ , \\
 \p_{\r_n}&=&\frac{V_{S^3}e^\a}{e^2N_0}\left[ \dot{\r}_n
               -2\c_0\a_n-n\tilde{\a}_n\right]
               +\frac{2}{e^2N_0}(\c_0^2-1)j_n\ , \\
 \p_{\s_n}&=&\frac{V_{S^3}e^\a}{3e^2N_0(n^2-1)}\left[\dot{\r}_n-\c_0\b_n\right]
             +\frac{1}{3e^2N_0(n^2-1)}(\c_0^2-1)k_n\ , \\
 \p_{\m_n}&=&\frac{V_{S^3}e^\a}{e^2N_0}\left[ \dot{\m}_n
             -2\dot{\c_0}d_n\right] \ , \\
 \p_{\n_n}&=&\frac{V_{S^3}e^\a(n^2-4)}{e^2N_0}\left[ \dot{\n}_n
             -\a_n-2\dot{\c}_0 c_n \right] \ , \\
 \p_{\x_n}&=&\frac{4V_{S^3}e^\a(n^2-4)}{e^2N_0(n^2-1)}
             \left[ \dot{\x}_n-{1 \over 6}\b_n -\dot{\c}_0b_n \right]\; .
 \label{momenta}
\eaee
Their quantized version can be used to generate our gauge
transformations~(\ref{trafo}) performing the commutator
$\d A_a=[A_a,w_n\ch_\a^{(n)}+v_n\ch_\b^{(n)}]$ with the harmonic
expansion~(\ref{haraa}) for $A_a$. The pure gravitational parts of
eq.~(\ref{ham_str}) and the momenta for the gravitational perturbations
can be found in ref.~\cite{hh}. We will not give the other
YM parts explicitly since we are mainly interested in the semi--classical
wave function in a region of superspace where the minisuperspace variables are
classical but the perturbations are quantum mechanical.
Therefore, it is enough in our approximation to solve first the
Euclidean background equations by setting the perturbations equal to zero and
then solve the equations of motion for the perturbations subject to the
Hartle-Hawking boundary condition by neglecting their back
reactions on the background. In this way, the wave function is given by
$\Psi=C \exp (-\hat I_E)$ where $C$ is a semi--classical prefactor and $\hat
I_E$
the extremal Euclidean action. $\hat I_E$ is given by $\hat I_E =
(1/2)p_n \pi _{p_n}$ where $p_n$ is a generic perturbation.

\section*{5. The Wave Function}

Let us follow the above program for the particular case of the EYM system which
we are considering here.
The background equations of motion for the minisuperspace degrees of freedom
$\a$, $\chi _0$ are given by
\be
e^{3\a} {d\over dt} \left( {{\dot \a}\over N_0}\right)
+{3\over 2}{e^{3\a}\over N_0}\dot \a ^2 +{1\over 2}N_0 e^\a
-{3\over 2} H^2 N_0 e^{3\a}
+{1\over 2 e^2}\left( {e^\a \over 2 N_0} \dot \chi _0 ^2
+{N_0\over e^\a}V \right)=0 \,,
\label{backa}
\ee
\be
 {e^\a \over N_0}{d\over dt}\left( {e^\a \over N_0} \dot{\c}_0 \right)
 +   V' =0 \,,
\label{backx}
\ee
while for $N_0$ we obtain the constraint
\be
{e^{3\a}\over 2}\left( {\dot\a ^2\over N_0^2 } + e^{-2\a} - H^2\right)
 - {1\over 2e^2N_0}\left( {e^\a \over 2 N_0} \dot \chi _0 ^2
 + {N_0\over e^\a}V\right)=0\,.
\label{n0}
\ee
Here $V'$ denotes the derivative of $V=1/2(1-\c_0^2)^2$
(cf.~eq.~(\ref{ym_pot0})) with respect to $\c_0$.
The perturbations of the gravitational and YM fields can be separated
into three types, namely tensor ($d_n$, $\m _n$),
 vector ($c_n$, $\n _n$, $j_n$, $\a _n$, $\rho _n$) and scalar
($a_n$, $b_n$, $\s _n$, $\g _n$, $\xi _n$, $\b _n$, $g_n$, $k_n$) modes.
In the equations of motion different types of
perturbations do not couple. This observation allows to
consider separately each type of perturbation and calculate their corresponding
contribution to the wave function.

 For the tensor perturbations $d_n$ and $\m_n$ we obtain the following
equations of motion
\bea
\ds {d\over dt} \left( \ds {e^{3\a}\over N_0} \dot d_n \right)
+ e^\a N_0 (n^2-1) d_n &=&
{N_0\over e^2 e^\a}\left[4\left({e^{2\a}\over 2N_0^2}\dot{\c}_0^2-V\right) d_n
+V'\m_n+2nU\tilde{\m}_n \right . \nnu \\
 &&\left . -2{e^{2\a}\over N_0^2}\dot{\c}_0\dot{\m}_n\right] \ ,
 \label{dn} \\
 {e^\a\over N_0}{d\over dt} \left( {e^\a\over N_0} \dot \m_n \right)
 +(n^2+1)\m_n &=&
 2{e^{2\a}\over N_0^2}\dot{\c}_0\dot{d}_n-2n\c_0\tilde{\m}_n+2nU\tilde{d}_n
 -V'd_n\ ,
\label{mun}
\eea
with the abbreviation $U=\c_0^2-1$.

Let us now consider the case of  the vector perturbations. To simplify the
calculation one can make a choice of gauge. For the gravitational field we
choose to proceed in the transverse-traceless gauge which corresponds for the
vector modes in $c_n=0$. For the YM field inspection of the
transformations (\ref{trafo}) shows that we are free to take $\rho _n=0$.
In this gauge the equations of motion and the constraints of the
remaining variables ($\n_n$, $j_n$, $\a_n$) are given by~:
\bea
{e^\a\over N_0}{d\over dt} \left({e^\a\over N_0}(\dot \n _n -\a_n)\right)
 +{1\over 2} n^2 \n_n &=& -\c _0 n \td \n _n
-{e^\a\over N_0^2} j_n\dot \c _0 \label{nun} \ , \\
\left[e^\a(n^2-4)+{2\over e^2 e^\a}V\right]j_n&=&
{1\over 4e^2}\left( 2V'\a_n +\dot \c_0 (n^2-4 )\n _n
+2nU\td \a _n\right) \label{jn} \ , \\
e^\a \left[ 2(n^2+2U) \a_n + 4n\c_0 \td\a_n \right]&=&
2(V' j_n +nU\tilde{j}_n) + e^\a(n^2-4)\dot\n_n\; .  \label{alphan}
\eea

In the case of the scalar modes the Lagrange multipliers are
$k_n$, $g_n$, $\b _n$ leaving five genuine variables. The transverse
traceless gauge for the gravitational degrees of freedom dictates for
the scalar modes $a_n=b_n=0$ and we will choose $\sigma _n=0$
for the YM modes. The remaining modes
($\g _n$, $\xi _n$, $\b _n$,  $g_n$, $k_n$) fulfill~:
\bea
 {e^\a\over N_0}{d\over dt}\left( {e^\a\over N_0}(\dot{\g}_n+
 {1 \over 3}\b_n) \right)+{2\over 3}(n^2+5+9U)\g_n+{4\over 3}(n^2-4)\x_n
 &=& {e^{2\a}\over N_0^2}\dot{\c}_0\dot{g}_n  - 2V' g_n\nnu  \\
  &&-{2e^\a\over 3N_0^2}\dot{\c}_0 k_n \label{gan}\ ,\\
 {e^\a\over N_0}{d\over dt}\left( {e^\a\over N_0}(\dot{\x}_n-
 {1\over 6}\b_n)\right)+{1\over 6}(n^2-1)(\g_n+2\x_n)
  &=& -{e^\a\over 6N_0^2}\dot{\c}_0 k_n\ ,
  \label{xin}
\eea
\bea
 \b_n&=&{1\over n^2-1+2U}\left[ 4(n^2-4)\dot{\x}_n-(n^2-1)\dot{\g}_n
        +(n^2-1)\dot{\c}_0g_n+e^{-\a}V' k_n \right] \nnu \ , \\
 k_n&=&{1\over e^2\dot{\a}e^\a}\left[{1\over 2}\dot{\c}_0(3\dot{\g}_n+\b_n)
       +{3N_0^2\over 2e^{2\a}}V'\g_n-{3 \over 2}\dot{\c}_0^2g_n\right]
       +3\dot{\a}e^\a g_n  \label{bekgn} \ , \\
 g_n&=&{1 \over e^2\dot{\a}e^{2\a}}\left[ \dot{\c}_0\g_n+2{n^2-4 \over n^2-1}
       \dot{\c}_0 \x_n+{1\over 2}{1\over n^2-1}V'\b_n-{2\over e^\a}
       {1\over n^2-1}V k_n\right]+{1\over\dot{\a}e^\a}{1\over n^2-1}k_n\; .
 \nnu
\eea
To calculate the wave function we need the value of the Euclidean action for
the classical solution which in our gauge is given by~:
\bea
{}~^T S_{cl}^{(n)}&=&{V_{S^3} \over 2}e^{2\a}\left(d_nd'_n+4\a'd_n^2\right)
            +{V_{S^3} \over 2e^2}\left( \m_n\m'_n-2\c'_0d_n\m_n\right)
\nnu \ ,\\
{}~^V S_{cl}^{(n)}&=&{V_{S^3}e^\a\over 2e^2 N_0}(n^2-4)(\dot{\n}_n-\a_n)\n_n
    \label{class_action} \ ,\\
{}~^S S_{cl}^{(n)}&=&{V_{S^3}e^\a\over e^2N_0}\left[{1\over 4}
                   (\dot{\g}_n+{1\over 3}\b_n-\dot{\c}_0 g_n)\g_n
                   +2{n^2-4\over n^2-1}(\dot{\x}_n-{1\over 6}\b_n)\x_n\right]
\; . \nnu
\eea
In the above expressions the prime denotes the derivative with respect
to the conformal time $\eta$ with ${d\eta \over dt}=-i{N_0\over e^\a}$.

We proceed to solve the above system of equations in the large $n$
approximation, i.~e.~we only consider the terms with the highest occurring
power
of $n$. After eliminating the auxiliary modes the equations decouple in this
approximation and can be written in terms of the gauge invariant variables
introduced in eq.~(\ref{inv}). They take the simple form $P''_n=n^2P_n$ for
$P_n = d_n,\m_n,R_n,\Gamma_n$.
Imposing the Hartle-Hawking boundary condition $P_n\rightarrow 0$ for
$\eta\rightarrow -\infty$ singles out the solutions $P_n\sim \exp (n\eta)$.
With the eqs.~(\ref{class_action}) this results in the wave function
\bea
{}~^T\Psi ^{(n)}&=& C \exp \left[ -{V_{S^3} \over2}e^{2\a}nd_n^2
                -{V_{S^3} \over 2e^2}n\m_n^2 \right ] \nnu\ , \\
{}~^V\Psi ^{(n)}&=& C^\prime \exp \left[ -{V_{S^3}\over 4e^2}
                n R_n^2\right] \label{wave_fct} \ ,\\
{}~^S\Psi ^{(n)}&=& C^{\prime \prime} \exp \left[ -{V_{S^3}\over 6e^2}
               n\Gamma_n^2 \right]\; .
\nnu
\eea
We see that
the perturbations start out in their ground state. In contrast
to the gravitational mode $d_n$ the frequencies corresponding to the YM
perturbations do not depend on the scale factor $e^\a$. This behavior is a
consequence of conformal invariance of YM theory. \\

A case of particular interest is when the classical field $\c_0$
obtains its minimum value $\c_0 = \pm 1$. Obviously this represents a
solution of the background equation of motion~(\ref{backx}) compatible with
the Hartle-Hawking boundary condition. As already discussed, the Lagrangian
in such a case can be explicitly expressed in terms of the gauge invariant
variables. An additional simplification occurs because $\c_0 = \pm 1$
corresponds to a symmetric field $A^{(0)}$ which is pure gauge as can be seen
{}from the eqs.~(\ref{F0_symm}), (\ref{F_symm}).
This causes the decoupling of gravitational and gauge
perturbations in the quadratic terms we are considering here. The Lagrangian
$L^n_{YM}=~^T\cl^n+~^V\cl^n+~^S\cl^n$ then takes the form~:
\bea
 ~^T\cl^n&=&{1\over 2}\left[ {e^\a \over N_0}\dot{\m}_n^2
            -{N_0\over e^\a}\left( (n^2+1)\m_n^2+2n\c_0\m_n\tilde{\m}_n
            \right) \right] \nnu \ , \\
 ~^V\cl^n&=&{e^\a \over 2N_0}\left[\dot{R}_n^2-4\c_0 \dot{R}_n A_n
            - 2n\dot{R}_n\tilde{A}_n + 2n^2A_n^2
            + 4\c_0 n A_n\tilde{A}_n \right] \nnu \\
            &&- {N_0\over 4e^\a}\left[ n^2R_n^2-2n\c_0R_n\tilde{R}_n\right]
             \label{sp_lagr} \ , \\
 ~^S\cl^n&=&{e^\a \over 2N_0}\left[{n^2-4 \over n^2-1}\dot{\Gamma}_n^2
            +{1\over 3(n^2-1)}\dot{S}_n^2-{2(n^2-4)\over 3(n^2-1)}B_n
            \dot{\Gamma}_n-{2\over 3(n^2-1)}\c_0 B_n\dot{S}_n
            +{1\over 6}B_n^2\right] \nnu \\
            && - {N_0\over 6e^\a}\left[ (n^2-4)\Gamma_n^2+S_n^2\right] \; .\nnu
\eea
Starting from these expressions it is easy to calculate the equations of motion
and eliminate the Lagrange multipliers $A_n$ and $B_n$. Solutions can be
found without any further approximation and we obtain the explicitly gauge
invariant wave function
\bea
{}~^T\Psi ^{(n)}&\sim& \exp \left[-{V_{S^3}\over 2e^2}(n\m_n^2
                     +\c_0\m_n\tilde{\m}_n) \right ] \nnu \ ,\\
{}~^V\Psi ^{(n)}&\sim&  \exp \left[ -{V_{S^3}\over 4e^2}
                (n R_n^2-2\c_0 R_n\td R_n)\right] \label{wave_fct_sp} \ ,\\
{}~^S\Psi ^{(n)}&\sim& \exp \left[ -{V_{S^3}\over 6e^2} (\Gamma_n,S_n)M^{(n)}
                   \left( \ba{c} \Gamma_n\\S_n \ea \right) \right] \ ,\nnu
\eea
with
\be
 M^{(n)}={1\over n(n^2-1)}\left( \ba{cc} (n^2+2)(n^2-4)&-\c_0 (n^2-4) \\
                                   -\c_0 (n^2-4)&n^2-2 \ea \right) \; .
\ee
As a consistency check one can apply the linear gauge constraints
$\ch_\a^{(n)}$, $\ch_\b^{(n)}$ and one finds that they annihilate
$~^V\Psi^{(n)}$ and $~^S\Psi^{(n)}$.\\

Finally we want to discuss some features of the general solutions
to the non--trivial systems of differential equations
which govern the classical evolution of the perturbations. The
solutions considered above have the drawback that they describe situations,
where the gravitational and gauge sectors are completely decoupled
as for the symmetric configuration. The separation
was due to the fact that the YM Lagrangian is scale invariant and
the only symmetric gravitational degree of freedom is that of the scale factor.
In the first case
the $n^2$--approximation is the first term of an approximation in rapidly
changing fluctuations and the decoupling appears as a consequence of the
diagonality of the kinetic terms \footnote{In fact this approximation
reduces to a large momentum one for large enough $n$ due to local
Lorentz invariance.}. The splitting in the
case where $\chi_0$ is fixed to the minimum of the potential is again a
property of the YM Lagrangian, together with the restriction to terms up to
second order in the non--symmetric fields:  From the general from of the
YM action each term containing
gravitational and gauge perturbations has to be supplemented by a component of
the background field strength. It is therefore multiplied  with a time
derivative of $\c_0$ or a term proportional to $\chi_0^2-1$ which
both vanish for $\c_0 = \pm 1$. This effect is an intrinsic
property of the symmetric YM potential and its knowledge in the
formalism used in this paper is crucial as opposed to the approach, where
the YM degrees of freedom are treated from the beginning
as scalar fields. Obviously the
above reasonings do not apply to expressions of higher
order in the perturbations and therefore there is no reason that the
decoupling survives even in these special cases.

To investigate the interactions between the gauge and gravitational
perturbations let us consider the next--to--leading--order
equations in a large $n$ approximation of the coupled system
describing the tensor modes, which are relevant for the discussion
of density perturbations. Invoking the ansatz
\be
\ba{llll} \label{eqdef}
d_n^\pm &\equiv 1/2(d_n^{(even)} \pm d_n^{(odd)}) \ , \qquad&
\mu_n^\pm &\equiv 1/2(\mu_n^{(even)} \pm \mu_n^{(odd)}) \\
&= e^{n\eta} D_n^\pm
&&= e^{n\eta} M_n^\pm \ ,
\ea
\ee
we find the following set of equations at order $n$:
\bea \label{eq001}
\frac{d}{d\eta} D_n^\pm - \tanh(\eta-\eta_a) D_n^\pm + \frac{H^2}{e^2}
\cosh(\eta-\eta_a)^2 f^\pm M^\pm_n &=& 0 \ ,\nn \\
\label{eq002}
\frac{d}{d\eta} M_n^\pm  + g^\pm M_n^\pm +
h^\pm D^\pm_n &=& 0 \ ,
\eea
where $d\eta = iN_0a^{-1}dt$. We recall at this point that the exponential
$e^{n\eta}$ in equation ~(\ref{eqdef}) is the leading order behaviour
found previously, while the prefactors $M_n^\pm$ and $D_n^\pm$ represent
the next--to--leading--order corrections. In eqs.~(\ref{eq001}) the
conformal mode $a$ has been replaced by the classical solution
\be
a^{-1}(\eta) = H \cosh(\eta-\eta_a)\ ,
\ee
where $\eta_a$ is defined by the
value of $a$ on the given three--surface, $a(0)=a_0$.
Inspection of
equations  (\ref{eq001}), (\ref{eq002}) reveals that the functions
$D_n^\pm$ and $M_n^\pm$ do not really depend on n; from now on we
simply denote them by $D_\pm, ~M_\pm$. Inspection of the same
equations shows that, unlike the leading order exponential, the
prefactors depend on the value of the field $\chi_0(\eta)$.
This field will be replaced by the classical solution given in
ref.~\cite{bm}, expressed through the functions $f,g$ and $h$
defined below:
\be
f^\pm = \pm(\chi_0^2 (\eta) -1)+\chi_0^\prime (\eta) \ , \qquad -g^\pm
= \pm \chi_0 (\eta)
\ , \qquad h^\pm =\pm(\chi_0^2 (\eta) -1)-\chi_0^\prime (\eta) \ .
\ee
There are four different solutions, determined by their boundary
values at $\eta = \infty$ and $\eta = 0$~:
\be \ba{lllcl}
I&:&\chi_0(\eta) = \coth(\eta-\eta_0) \ , &\chi_0 &<~-1 \ , \\
II&:&\chi_0(\eta) = \tanh(\eta-\eta_0) \ , &|\chi_0| &\leq~~1 \ , \\
III&:&\chi_0(\eta) = -\tanh(\eta-\eta_0) \ , &|\chi_0| &\leq~~1 \ , \\
IV&:&\chi_0(\eta) = -\coth(\eta-\eta_0) \ , &\chi_0 & >~~1 \ . \\
\ea \ee
On the other hand, $\eta_0$ expresses the boundary condition on
$\chi_0 (\eta)$ at $\eta=0~:~\chi_0 (0) = \chi_0$. For instance, for the
first of the above solutions, one gets: $ \coth(-\eta_0) = \chi_0 $; this
defines $\eta_0$ for this solution. We should remark that $\chi_0 (\eta)$
will remain in the region from which it started off at $\eta=0$.
The solutions of eqs.~(\ref{eq001}) corresponding to the four possible choices
for $\chi_0(\eta)$ are
\be\ba{lll}
\underline{I}:&&\\
D_+ &= &A_+\cosh(\eta-\eta_a) \ , \\
M_+ &= &A_+[\cosh(\eta_0-\eta_a) \sinh(\eta-\eta_0)^{-1}
+2\sinh(\eta_0-\eta_a)\cosh(\eta-\eta_0)]\\
&&+B_+\sinh(\eta-\eta_0) \ , \\
D_- &= &-\frac{H^2}{e^2}A_-
[\cosh(\eta_0-\eta_a)\cosh(\eta-\eta_a)\sinh(\eta-\eta_0)^{-2} \\
&&+2\ \sinh(\eta_0-\eta_a) \cosh(\eta-\eta_a) \coth(\eta-\eta_0)]
+ B_-\cosh(\eta-\eta_a)\ ,
\\
M_- &= &A_-\sinh(\eta-\eta_0)^{-1} \ ,
\ea\ee
\be\ba{lll}
\underline{II:}&&\\
D_+ &= &A_+\cosh(\eta-\eta_a)\ , \\
M_+ &= &A_+[2\cosh(\eta_0-\eta_a) \sinh(\eta-\eta_0)\\
&&-\sinh(\eta_0-\eta_a)\cosh(\eta-\eta_0)^{-1} ]
+B_+\cosh(\eta-\eta_0)\ , \\
D_- &= &-\frac{H^2}{e^2}
A_-[2\cosh(\eta_0-\eta_a)\cosh(\eta-\eta_a)\tanh(\eta-\eta_0) \\
&&-\sinh(\eta_0-\eta_a) \cosh(\eta-\eta_a) \cosh(\eta-\eta_0)^{-2} ]
+ B_-\cosh(\eta-\eta_a)\ , \\
M_- &= &A_-\cosh(\eta-\eta_0)^{-1} \ ,
\ea\ee
\be \ba{lll}
\underline{III}:&&\\
D_+ &= &\frac{H^2}{e^2}
A_+[2\cosh(\eta_0-\eta_a)\cosh(\eta-\eta_a)\tanh(\eta-\eta_0) \\
&&-\sinh(\eta_0-\eta_a) \cosh(\eta-\eta_a) \cosh(\eta-\eta_0)^{-2}]
+B_+\cosh(\eta-\eta_a)\ , \\
M_+ &= &A_+\cosh(\eta-\eta_0)^{-1} \ , \\
D_- &= &A_-\cosh(\eta-\eta_a) \ , \\
M_- &= &-A_-
[2\cosh(\eta_0-\eta_a)\sinh(\eta-\eta_0)\\
&&-\sinh(\eta_0-\eta_a) \cosh(\eta-\eta_0)^{-1}]
+B_-\cosh(\eta-\eta_0) \ ,
\ea\ee
\be \ba{lll}
\underline{IV}:&&\\
D_+ &= &\frac{H^2}{e^2}A_+
[\cosh(\eta_0-\eta_a) \cosh(\eta-\eta_a) \sinh(\eta-\eta_0)^{-2} \\
&&+2\sinh(\eta_0-\eta_a)\cosh(\eta-\eta_a)\coth(\eta-\eta_0)] + B_+
\cosh(\eta-\eta_a)\ , \\
M_+ &= &A_+\sinh(\eta-\eta_0)^{-1}\ ,  \\
D_- &= &A_-\cosh(\eta-\eta_a) \ , \\
M_- &= &-A_-[2\sinh(\eta_0-\eta_a)\cosh(\eta-\eta_0)\\
&&+\cosh(\eta_0-\eta_a)\sinh(\eta-\eta_0)^{-1} ]
+B_-\sinh(\eta-\eta_0) \ ,
\ea\ee
where $A_\pm$ and $B_\pm$ are integration constants. Note that
there are two types of relations between coefficients entering both
the gravitational and gauge perturbations: in the first case
the relative strength depends on the ratio $H/e$ while it
is universal in the second one.
Insofar as the physical properties of these solutions are concerned,
it is interesting to note that the interactions can
cause quite strong correlations between the two kinds of
fluctuations. For instance, if the ratio ${{M_\pm} \over {D_\pm}}$
is normalized to a fixed value for a given conformal time $\eta$,
it is predicted for any other value of $\eta$ and can differ
substantially from the initial choice. This might give rise to
interesting consistency conditions for cosmological models, if
this ratio is constrained for physical reasons at two
different times of the evolution.

Starting from the general expression~(\ref{class_action})
for the classical action and eliminating the derivatives via the
eqs.~(\ref{eq001},\ref{eq002}) one finds the tensor wave function
in the $O(n)$ approximation~:
\bea
 ~^T\Psi ^{(n)}&=&C\exp \left[ -{V_{S^3} \over2}e^{2\a}(n+3\a ')d_n^2\right. \\
               &&\left. -{V_{S^3} \over 2e^2}(n\m_n^2 + \c_0\m_n\tilde{\m}_n
                +2(1-\c_0^2)d_n\tilde{\m}_n - 2\c_0' d_n\m_n) \right ] \, .
\eea
As expected it shows a mixing between the gravitational and the YM
mode in the new subleading terms.

\section*{7. Conclusions and Outlook}

In this work we have extended the minisuperspace approximation treated in
previous works to include the full infinite-dimensional model of symmetric
gauge fields coupled to gravity.
A primary ansatz characterizing our approach is that we treated gauge fields,
such that the change of the gauge potential
due to the translation on $S^3$ can be compensated by a suitable gauge
transformation. $SO(N)$-symmetric gauge fields have been considered, however
the
extension to unitary, symplectic or exceptional groups is within reach;
in fact it should be made, if a realistic model is to be constructed within
this framework.

We restricted the gauge group to be just $SO(3)$,
to keep the formulae tractable. We feel that
this restriction does not hide too many characteristics of the system.
Most probably, a general $SO(N)$ group would have richer topological
structure, but we have not addressed this issue in this paper.
Probably working with different class of gauge groups would give rise
to essentially different phenomena. So, e.~g., the groups $SU(N)$ with
a conventional embedding of $SO(3)$ would impose the use of spinor harmonics.

Only a part of the superspace has been investigated, namely the portion
where the minisuperspace variables are classical, while the perturbations
are quantum mechanical. To investigate other regions, one should try to
solve the Wheeler-DeWitt equation in a better approximation. This would
allow taking into account the back reaction of the perturbations on
the background fields.
An interesting feature of our calculation has been the existence of
gauge invariant variables and the possibility to express the second
order Lagrangian in terms of them, illustrated in the main text for the
case $\chi_0 = \pm 1$.

We will present the detailed predictions of the wave function found above
in a future publication. For the time being we comment on some very basic
characteristics of this solution. The wave function has been found in
the regime of large n (the order of the corresponding harmonic).

As a first step,
we only kept the terms of order $n^2$; the wave function we
found shows that the perturbations start off in their ground state.
This means that approximating the inhomogeneous and anisotropic degrees
of freedom just to second order is well justified and describes the system
quite well in the relevant region of the superspace.
We also point out that the vector perturbations are genuine in our case
and not a gauge artifact, as is the case of the scalar fields.

When we proceed with the next-to-leading order, i.~e.~we also keep
terms of order $n$, we encounter new, quite interesting, phenomena.
Let us pay particular attention to the tensor perturbation modes, which
are e.~g.~relevant for the spectrum of relic gravitons in the universe.
We stress that, in contrast to models with scalar fields, we have a second
tensor mode in the theory stemming from the YM field. In the next to leading
order this mode couples to the gravity tensor perturbation often referred
to as the linear graviton.
A similiar behaviour shows up in the corresponding tensor mode wave function.
It is conceivable that these effects could have interesting cosmological
consequences.

One line of development of our work would be the inclusion of scalar and
fermionic fields and consider the full system at work.

Moreover, one may calculate wormhole effects using
our results. For example one can find the effective interactions between
gauge fields at large distances caused by the presence of wormholes.
This requires calculating the matrix element of the product of two
gauge field operators between the flat space vacuum state and wormhole
states. The latter one can be computed semiclassically starting from
our equations of motion for the perturbations and finding solutions which
fulfill the boundary conditions appropriate for wormhole states. \\

{\bf Acknowledgement}~This work was partially supported by the Deutsche
Forschungsgemeinschaft and the EC under contract no.~SC1-CT92-0789 and
the CEC Science Program no.~SC1-CT91-0729.

\renewcommand{\theequation}{\thesection.\arabic{equation}}
\newcommand{\sect}[1]{\section{#1}\setcounter{equation}{0}}

\section*{Appendices}
\appendix
\sect{Symmetric Fields}

Let us briefly summarize here how one can perform the harmonic analysis of
symmetric fields on homogeneous spaces. For the general method of constructing
symmetric fields we refer the reader to refs.~\cite{report} where a detailed
treatment is given. Let us just comment, that a field is called $S$ symmetric
on a space $M \times S/R$ if a symmetry transformation by an element of the
isometry group $S$ corresponds to a gauge transformation. This implies that
functions of the fields such as the energy-momentum tensor or
the Lagrangian density are independent of the coordinates of $S/R$ just because
they are gauge invariant. The requirement that transformations of the fields
under
the action of the symmetry group of $S/R$ are compensated by gauge
transformations, leads to certain constraints on the fields. The solution of
these constraints provides the theory on $M$. In particular, a
gauge field $A_{\sis M}$ on $M^{\sis D}$ splits into
$A_m$ on $M$ and $A_\alpha$ on $S/R$; $A_m$ behaves
as a scalar under $S$-transformations and lies in the
adjoint representation of the gauge group $G$.
The surviving gauge symmetry $H$ on $M$
is that subgroup of $G$ which commutes with $R$. In other words
the gauge group $H$ in one dimension is the
centralizer of $R$ in $G$, i.e.$~H=C_{\sis G}(R_{\sis G})$; $~ R_G$ is
the isomorphic image of $R$ in $G$.
The remaining components of the gauge field
 $A_\alpha$ become vectors under the coset space transformations.
The transformation properties of the fields $A_\a$
under $H$ can be found if
we express the adjoint irreducible representation of
$G$ in terms of $R_G\times H$
\beba
G &\supset & R_G\times H\ ,\\
{\rm adj}G &=& ({\rm adj}R,1)+(1,{\rm adj}H)+\sum_i (R_i,H_i)\,.
\label{2240}
\eaee
   and $S$ under $R$
\beba
S&\supset & R \ ,\\
{\rm adj}S &=& {\rm adj}R+\sum_i S_i \,.
\label{2241}
\eaee
   Then for every pair $R_i$, $S_i$ where $R_i$ and $S_i$ are identical
   irreducible representations of $R$, there remains a multiplet on $M$
   transforming under the representation $H_i$ of $H$. All other
    scalar fields
   vanish.

In order to perform the harmonic expansion of the fields on
$M\times S/R$
we have to find a suitable ground state configuration
around which to expand.
Since the symmetry of the vacuum $M\times S/R$ is the symmetry group of $M$
times $S$, we demand that the background configuration
$A_{\sis M}^{\sis B}$ should possess
these symmetries. Therefore $A_m^{\sis B}$ has to vanish while
$A_\a^{\sis B}$ has to be $S$-symmetric.
Such a configuration is given by
\be
A_\alpha ^{\sis B}=e_\alpha ^i J_i \ ,
\label{344}
\ee
where $\a = 1,\dots ,{\rm dim}S-{\rm dim} R$, $i=1,\dots , {\rm dim}R$,
$e_\a ^i$ are vielbeins of the coset space and $J_i$ are the generators of the
gauge group $G$ spanning the algebra of $R_{\sis G}$.
The background field $A_\alpha ^{\sis B}$ is
 also $S$-symmetric in the broader
sense i.e. including a suitable gauge
transformation. Indeed we obtain that
\be
\delta _{\sis A} A_\alpha ^{\sis B} =
 -(\partial _\alpha \Omega _{\sis A}^i) J_i
+ [ \Omega _{\sis A} ^i J_i , A_\alpha ^{\sis B}]\,,
\label{347}
\ee
where $\Omega _{\sis A}^i$ are the so called $R$-compensators,
and therefore the gauge transformation needed to compensate this
$S$-transformation is given by~:
\be
W_{\sis A}=-\Omega _{\sis A}^i J_i \,.
\label{348}
\ee

Let us now discuss the harmonic expansion of a
field on a coset space. It is well known that
the matrix elements $D_{pq}^{(m)}$ of the
inequivalent unitary irreducible
representations of $S$, indexed by $(m)$, serve as a complete
orthonormal basis into
which any function defined on $S$ can be expanded
(this is known as the Peter--Weyl theorem in group theory~\cite{218}).
Since $S$ is the isometry group of $S/R$, we can still use, in the case of the
coset space,
the matrix elements of the unitary irreducible representations of $S$ as
an orthonormal basis for the harmonic analysis, although these now have
to be properly restricted~\cite{17}. The harmonic
expansions of the fields $A_ m$, $A_a$  can be written as
\bea
A_m (x,y) &=& \sum _{m^\prime} \sum _{pq} \sqrt{ d_{m^\prime} \over
d_{\cd}} ~D_{pq}^{(m^\prime)}(L(y))~ a_m ~^{(m^\prime)}_{pq}(x)
\,,
\label{349}\\
A_a(x,y) &=& A_a^{\sis B}+\sum _m \sum _{pq} \sqrt{ d_m \over d_{\cd}}
 ~D_{pq}^{(m)}(L(y))~ \phi _a ~^{(m)}_{pq}(x) =
A_a^{\sis B} + {\overline A}_a\,,
\label{3410}
\eea
where $L(y)$ is a representative
element of each $R$-equivalence class, $d_m$ is the dimension of the
representation $m$ of $S$ and $d_\cd$ is the dimension of the $\cd$
representation of $R$.
As we will see, only those
irreducible representations of $S$ contribute to the
harmonic expansion of a field on $S/R$ in the representation $\cd$ of $R$,
which contain ${\cal D}$ in their
decomposition under $R$.
Furthermore, all the relevant
$D^{(m)}$'s contribute to the harmonic expansion only with those of
their
rows which correspond to the particular $R$-representation $\cd$.
In order to determine the relevant representations of $S$ which
contribute to the harmonic expansions (\ref{349}), (\ref{3410}), we should
demand that under a general $S$-transformation, which now includes the
gauge transformation given in eq.~(\ref{348}), the coefficients in these
expansions should transform according to some
irreducible representation of $S$~\cite{318}.
We find~:
\bea
D_{ps}^{(m)}(Q_i)\phi _a ~^{(m)}_{pq}+
f_{iab}\phi _b ~^{(m)}_{sq} - [J_i, \phi _a
{}~^{(m)}_{sq}]&=&0\,,
\label{3414} \\
D_{ps}^{(m)}(Q_i)a_m ~^{(m)}_{pq}-[J_i, a_m~^{(m)}_{sq}]&=&0
\,.
\label{3415}
\eea
Eqs.~(\ref{3414}), (\ref{3415}) specify which representations
 of $S$ are present
in the harmonic expansions of the corresponding fields given in
eqs.~(\ref{349}), (\ref{3410}). From the solution of eq.~(\ref{3414})
 we find that the
relevant representations of $S$ for the harmonic expansion of $A_a$ are
those which include in their restriction to $R$ the representations
of $R$ appearing in the cross product of $R_i$ and $S_i$ given in
eqs.~(\ref{2240}) and (\ref{2241}). Similarly, for the harmonic expansion of
$A_m$ one should include only those representations of $R$ which
appear in eq.~(\ref{2240}). All these $S$ representations
will contribute only with their rows which correspond to the
$R$ representation.
Furthermore,  the $S$-symmetric fields are the
$y$-independent ones, i.e. the first terms in the expansions
(\ref{349}), (\ref{3410}).

\sect{Harmonics}

Various versions of the harmonics on $S^3$ have appeared in the
literature~\cite{lif,camp,dow}. We
use the ones basically proposed in~\cite{camp},
which bear great similarity to the
harmonics used by Halliwell and Hawking, but their properties are much
more transparent and easy to derive from group theoretical considerations.
We refer the reader to~\cite{camp}
for a more complete account and present here some
basic steps, needed to adapt these considerations to our notation.
The manifold is $S^3$, or $SU(2)$, which is considered as  a coset space
of $SU(2)_L \otimes SU(2)_R$ over $SU(2)_{diag}$, where $SU(2)_{diag}$ is
defined as usual as the subgroup of $SU(2)_L \otimes SU(2)_R$ with
equal left and right rotations. The harmonics for a field on the above coset
corresponding to an angular momentum $J$, say, will
transform as the irreducible representations $\cd^{(J)}$
of $SU(2)_{diag}$. Such
representations may be sieved out of the tensor product representations
$D^{(m)}$ of the group $SU(2)_L \otimes SU(2)_R$
which contain $\cd^{(J)}$ upon
restriction to $SU(2)_{diag}$. The tensor representations are characterized
by the two angular momenta $j_L,j_R$ corresponding to $SU(2)_L, SU(2)_R$
respectively. For instance the scalar harmonics (corresponding to $J=0$) are
contained in
\be
(0,0), ({1 \over 2},{1 \over 2}), (1,1), ({3 \over 2}, {3 \over 2}),\ ...\ ,
\ee
while the vector harmonics ($J=1$) are contained in
\be
(0,1),(1,0),(1,1),({1 \over 2},{1 \over 2}),({1 \over 2}, {3 \over 2}),
\ ... \ ,
\label{vech}
\ee
As a matter of notation, we denote the harmonic characterized by $J$
constructed out of the pair $(j_L,j_R)$, by $D^{(j_L,j_R|J \xi)}_{pq}(L(y))$.
The indices $p$ and $q$ label matrix elements and $L(y)$ is the
representative element for the coset, at the point described by $y$ in
some local coordinate system. The additional index $\xi$ is used to
distinguish between the various $\cd^{(J)}$'s that may be
contained in $D^{(m)}$. In fact for the $SU(2)$ group only one $\cd^{(J)}$
appears for fixed $J$, so we completely drop this index in the sequel.
We also note that
the index $p$ is constrained
to  run only over the $2J+1$ values corresponding
to the relevant $\cd^{(J)}$. As explained in Appendix A,
any field on the coset space transforming
according to $\cd^{(J)}$ can be expanded as follows~:
\be
\psi_p^{(J)}(L(y))=\sum_{(j_L,j_R) \supset J}
\sum_q \psi_q^{(j_L,j_R|J)}
D^{(j_L,j_R|J)}_{pq}(L(y))\; .
\ee
The orthonormality relation reads~:
\be
\sum_p \int_{S^3} d \mu (y)
{}~[D^{(j_L,j_R|J)}_{pq}(L(y))]^*
{}~[D^{(j_L^\prime,j_R^\prime|J)}_{pq^\prime}(L(y))]
{}~=~{ {V_{S^3}(2J+1)} \over {(2j_L+1)(2j_R+1)} }
\delta^{j_L j_L^\prime} \delta^{j_R j_R^\prime}
\delta_{q q^\prime} \; .
\label{ornorm}
\ee
As the reader will notice, we define in the sequel the transverse harmonics
with a normalization equal to $\sqrt{{(2j_L+1)(2j_R+1)}
\over {(2J+1)} }$.
The motivation was simply to make the right hand side of eq.~(\ref{ornorm})
equal $V_{S^3}\delta^{j_L j_L^\prime} \delta^{j_R j_R^\prime} \delta_{q
q^\prime}$.
It is useful to define the quantities
\be
 n=j_L+j_R+1,\quad d=j_L-j_R,
\ee
in terms of which the Casimir operator of the $(j_L,j_R)$ representation of
$SO(4) \approx SU(2)_L \otimes SU(2)_R$ equals
\be
C_2(j_L,j_R) = n^2 + d^2 -1.
\ee
Two important results read~:
\beba
\nabla_a~D^{(j_L,j_R|J)}_{pq}(L(y))~
&=&~-[T_a^{(j_L,j_R)}]_{pr}~D^{(j_L,j_R|J)}_{rq}(L(y))\ , \\
\nabla_a \nabla^a
{}~D^{(j_L,j_R|J)}_{pq}(L(y))~
&=&~[C_2(\cd^{(J)})-C_2(j_L,j_R)]~D^{(j_L,j_R|J)}_{pq}(L(y))\\
&=&~[J(J+1)-(n^2+d^2-1)]~D^{(j_L,j_R|J)}_{pq}(L(y)) \ .
\label{thmb4}
\eaee
In the above the $T_a^{(j_L,j_R)}$ are the part of the reductive
decomposition $(T_a^{(j_L,j_R)}$,$T_i^{(j_L,j_R)})$ of the generators of
$(j_L,j_R)$, which corresponds to $SU(2)_{diag}$; $C_2(\cd^{(J)})=J(J+1)$
is the Casimir operator for the
representation $J$ defining the harmonic. These important results mean
that any relation involving differentials on $S^3$ reduces to mere algebra.
Two more identities are
needed to find the properties of the harmonics. These are the
commutation relations for covariant derivatives acting on vectors and
tensors~:
\beba
(\nabla_a \nabla_c - \nabla_c \nabla_a)V_b~&=~&V_dR^d_{bac},\\
(\nabla_a \nabla_b - \nabla_b \nabla_a)H_{cd}~&=~&H_{ed}R^e_{cab}+
H_{ce}R^e_{dab},
\eaee
\be
R_{abcd}~=~\d_{bc} \d_{ad} - \d_{ac} \d_{bd}.
\ee
In the
following we give explicitly some properties of the scalar, vector and
tensor harmonics, which are the only relevant ones for the case $G=SO(N)$,
whose special case $N=3$ is considered in this work.

\noindent{\bf 1) Scalar harmonics}

These are scalar eigenfunctions of the Laplacian. It is convenient to
define them with the following normalization~:
\be
Q_q^{(n)}(L(y))={n^2} ~D^{(j_L,j_R|0)}_{0q}(L(y))\; .
\ee
The first index on $D^{(j_L,j_R|0)}_{0q}(L(y))$ takes the value
corresponding to the identity representation, so there is no point in
keeping it in the notation for the harmonic $Q^{(n)}_q$. We also use the
index $(n)$ defined above to characterize the scalar harmonics, rather
than the pair $(j_L,j_R)$.
In fact for the scalar harmonics, $j_L=j_R \equiv j$,
so the other combination, namely $d$, vanishes in this case. The quantity $j$
runs over integer and half-odd-integer values.
For the normalization introduced above, it is easy to show that
\be
\int_{S^3} d \mu (y)
{}~[Q^{(n)}_q(L(y))]^*~[Q^{(n^\prime)}_{q^\prime}(L(y))]
{}~=~V_{S^3} \delta^{n n^\prime} \delta_{q q^\prime} \; .
\ee
The eigenvalue of the Laplacian is found from the difference of the
Casimir operators (eq.~(\ref{thmb4}))~:
\be
\nabla_a \nabla^a~Q^{(n)}_q(L(y))
{}~=~-(n^2-1)~Q^{(n)}_q(L(y))\; .
\ee

\noindent{\bf 2) Vector harmonics}

The vector harmonics on a simply connected coset space without boundary
can be uniquely decomposed into a sum of the ``transverse vector
harmonics", characterized by the vanishing of their covariant divergence
and the ``longitudinal vector harmonics", given by the  gradient of some
(scalar) function. This corresponds to the fact, obvious from eq.~(\ref{vech}),
that the quantity $d$  may take three values, namely $-1,0$, and $+1$.
Thus the vector harmonics
split into categories, according to the value of $d$. The ones corresponding
to $d=-1~(d=+1)$ are called odd (even) transverse vector harmonics; we
denote them by $S^{(n-)}_{a|q}(L(y)),~S^{(n+)}_{a|q}(L(y))$ respectively,
while for $d~=~0$ we get the longitudinal vector harmonics,
$P^{(n)}_{a|q}(L(y))$. The index $a$ runs from 1 to 3.
We note that the terms ``odd" (``even") derive from the behavior of the
corresponding harmonics under a parity transformation~\cite{lif}.
Let us now be more specific and give the exact definitions of the various
harmonics~:
\be
S^{(n \pm)}_{a|q}(L(y))~ =~ \sqrt{{n^2-1} \over {3} }
D^{(j \pm 1,j|1)}_{aq}(L(y)) \; .
\ee
On the other hand, since the scalar harmonics, previously defined, span
the space of scalar functions on the coset space, the longitudinal vector
harmonic may be chosen to be the gradient of a scalar harmonic~:
\be
P_{a|q}^{(n)}(L(y))~ =~ {1 \over {n^2-1}} \nabla_a Q^{(n)}_q (L(y)).
\ee
Using the above definitions, we obtain the following properties of the
vector harmonics~:
\beba
\sum_a \int_{S^3} d \mu (y)
{}~[S^{(n \pm)}_{a|q}(L(y))]^*~[S^{(n^\prime \pm)}_{a|q^\prime}(L(y))]
{}~&=~& V_{S^3}\delta^{n n^\prime} \delta_{q q^\prime}\ ,\\
\sum_a \int_{S^3} d \mu (y)
{}~[S^{(n \pm)}_{a|q}(L(y))]^*~[S^{(n^\prime \mp)}_{a|q^\prime}(L(y))]~&=~&0\
,\\
\sum_a \int_{S^3} d \mu (y)
{}~[P^{(n)}_{a|q}(L(y))]^*~[P^{(n^\prime)}_{a|q^\prime}(L(y))]
{}~&=~&{\displaystyle V_{S^3} \over \displaystyle{n^2-1} }
 \delta^{n n^\prime} \delta_{q q^\prime}\ ,
\eaee
\beba
\nabla^a S^{(n \pm)}_{a|q}(L(y))~&=~&0\ ,\\
\nabla^a P^{(n)}_{a|q}(L(y))~&=~&-Q^{(n)}_q(L(y))\ ,\\
\ep_{acb}\nabla _a  S^{(n \pm)}_{b|q}(L(y)) ~&=~& n S^{(n \mp)}_{c|q}(L(y))\ ,
\eaee
\beba
\nabla_a \nabla^a~S^{(n \pm)}_{b|q}(L(y))
{}~&=~&-(n^2-2)~S^{(n \pm)}_{b|q}(L(y))\ ,\\
\nabla_a \nabla^a~P^{(n)}_{b|q}(L(y))
{}~&=~&-(n^2-3)~P^{(n)}_{b|q}(L(y)) \; .
\eaee

\noindent{\bf 3) Tensor harmonics}

As in the previous case, also the tensor harmonics can be classified
into several kinds, according to the value of $d \equiv j_L -j_R$. This
quantity may now take the values $\pm 2, \pm 1, 0$.

The harmonics with $d=-2 ~(d=+2)$ are called odd (even)
transverse tensor harmonics. To get them from the $D^{(j_L,j_R|2)}_{pq}$, we
represent the index $p$ (which may take $5$ values)  by a pair of indices
$a_1 a_2$, each running from 1 to 3. Then the definition is
\be
G_{a_1 a_2|q}^{(n \pm)}(L(y)) \equiv \sqrt{ {n^2-4} \over {5 } }
\sum_{q_1,q_2}
P(a_1 a_2|q_1 q_2) D^{(j \pm 2,j|2)}_{(q_1 q_2)q}(L(y)),
\ee
where the definition of the projection operator $P(a_1 a_2|q_1 q_2)$
reads~:
\be
P(a_1 a_2|q_1 q_2) \equiv [{1 \over 2} (\delta_{q_1 a_1} \delta_{q_2 a_2}
+ \delta_{q_2 a_1} \delta_{q_1 a_2})
-{1 \over 3} \delta_{q_1 q_2} \delta_{a_1 a_2}].
\ee
This operator projects out the symmetric and traceless part of the
expression it acts on.
Thus the harmonics $G_{a_1 a_2|q}^{(n \pm)}(L(y))$ are symmetric,
traceless and transverse.
We find the following additional
properties of the transverse tensor harmonics~:
\beba
\sum_{a_1,a_2} \int d\mu(y) [G_{a_1 a_2|q}^{(n \pm)}(L(y))]^*
[G_{a_1 a_2|q^\prime}^{(n^\prime \pm)}(L(y))]
{}~&=~&V_{S^3}\delta^{n n^\prime} \delta_{q q^\prime}\ ,\\
\sum_{a_1,a_2} \int d\mu(y) [G_{a_1 a_2|q}^{(n \pm)}(L(y))]^*
[G_{a_1 a_2|q^\prime}^{(n^\prime \mp)}(L(y))]~&=~&0\ ,
\eaee
\beba
\nabla_a \nabla^a G^{(n \pm)}_{a_1 a_2|q}(L(y))
{}~& =~& -(n^2-3) G^{(n \pm)}_{a_1 a_2|q}(L(y))\ ,\\
\nabla^{a_1} G_{a_1 a_2|q}^{(n \pm)}(L(y))~&=~&0\ ,\\
\delta_{a_1 a_2} G_{a_1 a_2|q}^{(n \pm)}(L(y))~&=~&0\ ,\\
\ep_{a_1 a_2 a_3} \nabla _{a_1} G^{(n \pm)}_{a_4 a_3|q}(L(y)) ~&=~&
n G^{(n \mp)}_{a_4 a_2|q}(L(y))\; .
\eaee

The remaining types of tensor harmonics can be expressed in the form~:
\be
\sum_{q_1, q_2}
P(a_1 a_2|q_1 q_2)
\nabla_{q_1} D^{(j_L,j_R|1)}_{q_2|q}(L(y)).
\label{long}
\ee
Basically they are covariant derivatives of harmonics with smaller
``magnetic quantum numbers" $d$, properly symmetrized.
We recall that any vector harmonic is a linear combination of the transverse
and longitudinal vector harmonics. Thus these remaining harmonics are
linear combinations of~:

1) The even and odd "longitudinal-transverse"
harmonics, with $|d| \equiv |j_L-j_R|=1$. Eq. (\ref{long}) becomes~:
\be
S_{a_1 a_2|q}^{(n \pm)}(L(y))~=~{1 \over 2} [\nabla_{a_1}
S_{a_2}^{(n\pm)}(L(y))
+\nabla_{a_2} S_{a_1}^{(n \pm)}(L(y))]
\ee
and

2) the "longitudinal-longitudinal" tensor harmonics, with $d \equiv j_l-j_R=0$.
Now eq.~(\ref{long}) becomes
\be
P_{a_1 a_2|q}^{(n)}(L(y))~=~{1 \over 2} [ \nabla_{a_1} P_{a_2 |q}^{(n)}(L(y))
+\nabla_{a_2} P_{a_1 |q}^{(n)}(L(y))]-{1 \over 3} \delta_{a_1 a_2}
\nabla^a P_{a|q}^{(n)}(L(y)) \ ,
\ee
respectively. From the definition of $P^{(n)}_{a|q}(L(y))$ one can prove that
\be
P_{a_1 a_2|q}^{(n)}(L(y)) = {1 \over {n^2-1}} \nabla_{a_1} \nabla_{a_2}
Q^{(n)}_q(L(y))
+{1 \over 3} \delta_{a_1 a_2} Q^{(n)}_q(L(y)),
\ee
which is the definition of this quantity used by Halliwell and Hawking.

In the following we gather some properties of these harmonics.
\beba
\sum_{a_1,a_2} \int d\mu(y) [S_{a_1 a_2|q}^{(n \pm)}(L(y))]^*
[S_{a_1 a_2|q^\prime}^{(n^\prime \pm)}(L(y))]
{}~&=~&V_{S^3}{\displaystyle (n^2-4) \over \displaystyle2}
 \delta^{n n^\prime} \delta_{q q^\prime}\ , \\
\sum_{a_1,a_2} \int d\mu(y) [S_{a_1 a_2|q}^{(n \pm)}(L(y))]^*
[S_{a_1 a_2|q^\prime}^{(n^\prime \mp)}(L(y))]~&=~&0\ , \\
\sum_{a_1,a_2} \int d\mu(y) [P_{a_1 a_2|q}^{(n)}(L(y))]^*
[P_{a_1 a_2|q^\prime}^{(n^\prime)}(L(y))]
{}~&=~&V_{S^3}{{ \displaystyle 2(n^2-4)} \over {\displaystyle 3(n^2-1)} }
\delta^{n n^\prime} \delta_{q q^\prime}\ ,
\eaee
\beba
\nabla^{a_1} S^{(n \pm)}_{a_1 a_2|q}(L(y))~&
=~& -(n^2-4) S_{a_2|q}^{(n \pm)}(L(y))\ ,\\
\nabla^{a_1} \nabla^{a_2} S^{(n \pm)}_{a_1 a_2|q}(L(y))~& =~& 0\ ,\\
\nabla^a \nabla_a S^{(n \pm)}_{a_1 a_2}(L(y))~&
=~&-(n^2-6) S^{(n \pm)}_{a_1 a_2|q}(L(y))\ ,\\
\nabla^a \nabla_a P^{(n)}_{a_1 a_2|q}(L(y))~&
=~&-(n^2-7) P^{(n)}_{a_1 a_2|q}(L(y))\ ,\\
\delta_{a_1 a_2} P_{a_1 a_2|q}^{(n)}(L(y))~&=~& 0\ ,\\
\nabla^{a_1} P_{a_1 a_2|q}^{(n)}(L(y))~&
=~& -{2 \over 3} (n^2-4) P_{a_2|q}^{(n)}(L(y))\ ,\\
\nabla^{a_1} \nabla^{a_2} P_{a_1 a_2|q}^{(n)}(L(y))~&=~&~
{2 \over 3}(n^2-4) Q^{(n)}_q(L(y))\ .
\eaee

\end{document}